\RequirePackage{etoolbox}

\makeatletter
\newcommand{\iftogglehard}[1]{%
	\ifcsdef{etb@tgl@#1}
		{\csname etb@tgl@#1\endcsname\iftrue\iffalse}
		{\etb@noglobal\etb@err@notoggle{#1}\iffalse}%
}
\makeatother

\newtoggle{draft}			

\newtoggle{showoverflow}		
\toggletrue{showoverflow}

\newtoggle{allowtodo}		
\toggletrue{allowtodo}

\newtoggle{showtodo}		
\toggletrue{showtodo}

\newtoggle{inlinetodo}		
\toggletrue{inlinetodo}

\newtoggle{printversion}		

\newtoggle{conference}		

\newtoggle{llncs}			

\newtoggle{twocols}			

\newtoggle{anon}			

\iftoggle{conference}{
    \iftoggle{twocols}{
        \documentclass[letterpaper,twocolumn,10pt]{article}
    }{
		\documentclass[envcountsect]{llncs}
    }
}{
	\documentclass{article}
}

\newtoggle{beamer}
\ifcsdef{frametitle}{\toggletrue{beamer}}{}

\usepackage[utf8]{inputenc}
\usepackage[T1]{fontenc}

\usepackage{pdfpages} 

\usepackage{amsmath,amsfonts,amssymb}

\nottoggle{beamer}{
  \usepackage{enumitem}
}{}
\usepackage{framed}

\usepackage{caption}
\usepackage{subcaption}
\usepackage{dblfloatfix}

\usepackage{csquotes}

\usepackage{multicol}
\usepackage{boxedminipage}

\nottoggle{beamer}{
  \usepackage{footnote}
  \makesavenoteenv{tabular}
  \makesavenoteenv{table}
  \makesavenoteenv{table*}
}{}
\usepackage{multirow}

\usepackage{bm}

\usepackage{breakcites}

\usepackage{hyphenat}
\usepackage{xparse}

\nottoggle{beamer}{
  \iftoggle{allowtodo}{
    \iftoggle{showtodo}{
      \iftoggle{inlinetodo}{
        \newcommand{\todo}[1]{\textcolor{red}{[{\footnotesize {\bf TODO:} { {#1}}}]}}
      }{
        \usepackage[textsize=small]{todonotes}
      }
      \usepackage{marvosym}

    }{
      \usepackage[disable]{todonotes}
      \renewcommand{\todo}[1]{}
      \newcommand{brice}[1]{}
    }
    \usepackage{xspace}
    \expandafter\apptocmd\csname\string\todo\endcsname{\xspace}{}{}
  }{
  }
}{}

\usepackage{algorithm}
\usepackage{algorithmicx}
\usepackage[noend]{algpseudocode}

\usepackage{tabularx}

\algnewcommand\algorithmicto{\textbf{to}}
\algrenewtext{For}[3]%
  {\algorithmicfor\ $#1 = #2$ \algorithmicto\ $#3$ \algorithmicdo}

\algblockdefx[IFRED]{IfRed}{EndIfRed}
	[1]{\red{\algorithmicif\ {#1}}}
   	[0]{\red{\algorithmicend\ \algorithmicif}}

\algcblockx{IFRED}{ElseRed}{EndIfRed}
	[0]{\red{\algorithmicelse\ }}
	[0]{\red{\algorithmicend\ \algorithmicif}}
   
\algblockdefx[FORALLRED]{ForAllRed}{EndForRed}
   [1]{\red{\algorithmicforall\ {#1}}}
   [0]{\red{\algorithmicend\ \algorithmicfor}}

\algloopdefx[SMALLIF]{SmallIf}
	[1]{\algorithmicif\ {#1}}   
   
\algloopdefx[SMALLIFRED]{SmallIfRed}
	[1]{\red{\algorithmicif\ {#1}}}  
	 
\algblockdefx[GAME]{GameBlock}{EndGameBlock}
   [1]{{#1} Game:}
   [0]{}
\algtext*{EndGameBlock} 

\algdef{SE}[DOWHILE]{Do}{doWhile}{\algorithmicdo}[1]{\algorithmicwhile\ #1}%

\algrenewcommand\algorithmicindent{1.0em}%
\algrenewcommand\alglinenumber[1]{\scriptsize #1:}
\algrenewcommand{\algorithmiccomment}[1]{\hfill\hspace{.2cm} $\triangleright$\ \textit{#1}}

\usepackage{float}

\newfloat{protocol}{tbph}{lop}
\floatname{protocol}{Protocol}

\usepackage[lambda,advantage,adversary,landau,sets,notions,logic,ff,primitives,events,asymptotics,keys]{cryptocode}

\usepackage{tikz}
\usetikzlibrary{shapes,positioning,calc}
\usetikzlibrary{arrows}
\usetikzlibrary{fit}

\usepackage{import}

\usepackage[binary-units,mode=text]{siunitx}

\usepackage{amsthm}

\definecolor{blue_link}{RGB}{0,0,150}
\nottoggle{beamer}{

    \usepackage[pageanchor=false,pdfpagelabels=true]{hyperref}

  \hypersetup{
    colorlinks     = \iftoggle{printversion}{false}{true},
    linkcolor      = blue_link,
    citecolor      = blue_link,
    urlcolor       = blue_link,
    bookmarksdepth = 3,
    pdfborder = {0 0 0}
  }
}

\makeatletter
\AtBeginDocument{
  \hypersetup{
    pdftitle  = {\@title},
    pdfauthor = {\@author}
  }
}
\makeatother

\nottoggle{beamer}{
  \usepackage{cleveref}
  \crefname{lemma}{lemma}{lemmas}
  \Crefname{lemma}{Lemma}{Lemmas}
  \crefname{claim}{claim}{claims}
  \Crefname{claim}{Claim}{Claims}
}{}

\nottoggle{beamer}{

    \iftoggle{conference}{

      \spnewtheorem*{theorem*}{Theorem}{\bfseries}{\itshape}
      \spnewtheorem*{proposition*}{Proposition}{\bfseries}{\itshape}

    }{

      \newtheorem{theorem}{Theorem}

      \newtheorem{lemma}[theorem]{Lemma}

      \newtheorem{definition}{Definition}[section]

      \makeatletter
      \newtheorem*{rep@theorem}{\rep@title}
      \newcommand{\newreptheorem}[2]{%
        \newenvironment{rep#1}[1]{%
          \def\rep@title{#2 \ref{##1}}%
          \begin{rep@theorem}}%
            {\end{rep@theorem}}}
      \makeatother

      \newreptheorem{theorem}{Theorem}
      \newreptheorem{proposition}{Proposition}
    }
}{}

\newcommand{\RI}[2]{\ensuremath{[#1,#2]_{\R}}}     
\newcommand{\RIext}[2]{\ensuremath{(#1,#2)_{\R}}}  
\newcommand{\Int}[1]{\ensuremath{\{1,\dots,#1\}}}

\newcommand{\conc}{\ensuremath{\mid\mid}}

\newcommand{\para}[1]{\medskip{\bf #1.}}

\makeatletter
\newcommand{\multiline}[1]{%
  \begin{tabularx}{\dimexpr\linewidth-\ALG@thistlm}[t]{@{}X@{}}
    #1
  \end{tabularx}
}
\makeatother

\renewcommand{\mod}{\;\mathsf{mod}\;}

\renewcommand\dots{\makebox[1em][c]{.\hfil.\hfil.}}

\DeclareDocumentCommand\polylog{ g }{
\ensuremath{\mathsf{polylog}
\IfNoValueF {#1}{(#1)}
}}

\def\bit{\{0,1\}}

\def\N{\mathbb{N}}
\def\R{\mathbb{R}}

\newcommand{\eps}{\varepsilon}

\newcommand{\ie}{\emph{i.e.}~}

\newcommand{\bigo}{\mathcal{O}}

\newcommand{\ceil}[1]{\left\lceil #1 \right\rceil}
\DeclarePairedDelimiter\abs{\lvert}{\rvert}

\DeclareMathOperator*{\Exp}{\mathbb{E}}

\newcommand{\Enc}{\enc}
\newcommand{\Dec}{\dec}

\newcommand{\PRF}{\ensuremath{\mathsf{PRF}}}
\newcommand{\Hash}{\ensuremath{\mathsf{H}}}

\newcommand{\KeyGen}{\pcalgostyle{KeyGen}}

\newcommand{\bigOtilde}[1]{\ensuremath{\widetilde{\mathcal{O}}\left({#1}\right)}}

\newcommand{\INDCPA}{\mathsf{IND}\pcmathhyphen{}\mathsf{CPA}}

\usepackage{pifont}
\mathchardef\mhyphen ="2D

\newcommand{\red}[1]{\textcolor{red}{#1}}

\newcommand{\RN}[1]{%
  \textup{\uppercase\expandafter{\romannumeral#1}}%
}

\newcommand{\SSE}{\mathsf{SSE}}

\newcommand{\new}{^\mathsf{new}}

\newcommand{\encr}{^\mathsf{enc}}
\newcommand{\addlist}{\cup}

\newcommand{\FAIL}{\mathsf{FAIL}}

\newcommand{\SSEReal}{\textup{\textsc{SSEReal}}}
\newcommand{\SSEIdeal}{\textup{\textsc{SSEIdeal}}}
\newcommand{\adaptive}{\mathsf{adp}}
\newcommand{\selective}{\mathsf{sel}}

\newcommand{\Sim}{\textsf{Sim}}

\newcommand{\DB}{\mathsf{DB}}

\renewcommand{\ind}{\mathsf{id}}

\newcommand{\EDB}{\mathsf{EDB}}

\newcommand{\LR}{\mathcal{L}_{\mathsf{len}\text{-}\mathsf{rev}}}
\newcommand{\lr}{^{\mathsf{len}\text{-}\mathsf{rev}}}
\newcommand{\LH}{\mathcal{L}_{\mathsf{len}\text{-}\mathsf{hid}}}
\newcommand{\lh}{^{\mathsf{len}\text{-}\mathsf{hid}}}
\renewcommand{\L}{\mathcal{L}}
\newcommand{\Stp}{\mathsf{Stp}}
\newcommand{\Srch}{\mathsf{Srch}}

\newcommand{\Updt}{\mathsf{Updt}}

\newcommand{\Hist}{\mathsf{Hist}}

\newcommand{\qpat}{\mathsf{qp}}
\newcommand{\spat}{\mathsf{sp}}

\newcommand{\RdPat}{\mathsf{RdPat}}
\newcommand{\PgPat}{\mathsf{PgPat}}

\newcommand{\Setup}{\mathsf{Setup}}
\newcommand{\Search}{\mathsf{Search}}

\newcommand{\Update}{\mathsf{Update}}

\newcommand{\op}{\mathsf{op}}
\newcommand{\add}{\mathsf{add}}
\newcommand{\del}{\mathsf{del}}

\newcommand{\opin}{\mathsf{in}}

\newcommand{\KPRF}{\mathsf{K}_{\PRF}}
\newcommand{\KENC}{\mathsf{K}_{\Enc}}
\newcommand{\K}{\mathsf{K}}
\newcommand{\TLen}{T_\mathsf{len}}
\newcommand{\TPos}{T_\mathsf{pos}}

\newcommand{\TFull}{T_\mathsf{full}}

\newcommand{\Stash}{\mathsf{Stash}}

\newcommand{\tethysoram}{\mathsf{OramTethys}}
\newcommand{\uselessSSE}{\mathsf{UncondSSE}}
\newcommand{\pageSSE}{\mathsf{LayeredSSE}}
\newcommand{\Tethys}{\mathsf{Tethys}}

\newcommand{\localORAM}{\mathsf{LocORAM}}

\newcommand{\ORAMInit}{\mathsf{Initialize}_\mathsf{ORAM}}
\newcommand{\ORAMAccess}{\mathsf{Access}_\mathsf{ORAM}}

\newcommand{\Binpack}{\mathsf{Binpack}}
\newcommand{\OblSort}{\mathsf{OblSort}}

\newcommand{\Mem}{\mathsf{M}}
\newcommand{\EMem}{\mathsf{EM}}

\newcommand{\orambs}{\beta}

\newcommand{\cnt}{\mathsf{cnt}}
\newcommand{\T}{\mathsf{T}}
\newcommand{\found}{\mathsf{fnd}}
\newcommand{\indA}{\mathsf{index}}

\newcommand{\onechoice}{\textsf{1C}}
\newcommand{\twochoice}{\textsf{2C}}
\newcommand{\layeredchoice}{\textsf{L2C}}

\newcommand{\InsertBall}{\mathsf{InsertBall}}
\newcommand{\UpdateBall}{\mathsf{UpdateBall}}
\newcommand{\newball}{b_\mathsf{new}}
\newcommand{\newweight}{w_\mathsf{new}}
\newcommand{\oldweight}{w_\mathsf{old}}
\newcommand{\oldball}{b_\mathsf{old}}
\newcommand{\leftover}{residual\xspace}
\newcommand{\maxweight}{w_\mathsf{max}}

\newcommand{\onechoicealloc}{\ensuremath{\mathsf{1C\mhyphen Alloc}}}
\newcommand{\Fetch}{\ensuremath{\mathsf{Fetch}}}
\newcommand{\Append}{\ensuremath{\mathsf{Add}}}
\newcommand{\peff}{\ensuremath{\mathsf{PE\mhyphen SSE}}\xspace}
\newcommand{\nlevel}{\ensuremath{n_{\it level}}}
\newcommand{\clip}{\ensuremath{\mathsf{clip}}}
\newcommand{\clippedOSSE}{\ensuremath{\mathsf{ClipOSSE}}}
\newcommand{\Send}{\ensuremath{\mathbf{send}}\xspace}

\DeclareDocumentCommand\local{ g }{
\ensuremath{\mathsf{Local}
\IfNoValueF {#1}{[#1]}
}}

\newcommand{\ash}{Asharov et al.\xspace}

\newcommand{\GLT}{Generic Local Transform\xspace}

\nottoggle{conference}{
	\setcounter{tocdepth}{2}
	\usepackage{fullpage}
	\newtheorem{remark}{Remark}
}



\iftogglehard{draft}
\usepackage{eso-pic}

\makeatletter
\newlength\@tempdim@x
\newlength\@tempdim@y
\newcommand\AtCenterLeft[3]{%
\begingroup
\@tempdim@x=0cm
\@tempdim@y=.5\paperheight
\advance\@tempdim@x#1
\advance\@tempdim@y#2
\put(\LenToUnit{\@tempdim@x},\LenToUnit{\@tempdim@y}){#3}%
\endgroup
}

\AddToShipoutPicture{%
\AtCenterLeft{1cm}{0cm}{
\rotatebox[origin=c]{90}{{\Large UNDER SUBMISSION \raisebox{0.1ex}{--} DO NOT DISTRIBUTE}}}
}
\makeatother
\fi 

\iftoggle{llncs}{
	\pagestyle{plain} 
}{}

\begin{document}

\title{Dynamic Local Searchable Symmetric Encryption}

\nottoggle{anon}{
\iftoggle{conference}{
\author{
	Angèle Bossuat\inst{1}
	\and
	Raphael Bost\inst{2,}\thanks{ The views and conclusions contained herein are those of the author and should not be interpreted as necessarily representing the official policies or endorsements, either expressed or implied, of the DGA or the French Government.}
	\and
	Pierre-Alain Fouque\inst{1}
	\and
	Brice Minaud\inst{3,4}
	\and
	Michael Reichle\inst{3,4}
}
\institute{Quarkslab, and Université de Rennes 1, Rennes, France.
\and
Direction Générale de l'Armement, Paris, France.
\and
Inria, Paris, France.
\and
École Normale Supérieure, CNRS, PSL, Paris, France.}
}{
\author{Brice Minaud$^{1,2}$, Michael Reichle$^{1,2}$}
\date{
\normalsize
$^1$ Inria, Paris, France\\
$^2$ \'Ecole Normale Sup\'erieure, CNRS, PSL University, France}
}
}

\iftoggle{conference}{
\date{}
\institute{}
}

\maketitle 


\begin{abstract}
In this article, we tackle for the first time the problem of \emph{dynamic} memory-efficient Searchable Symmetric Encryption (SSE). In the term ``memory-efficient'' SSE, we encompass both the goals of \emph{local} SSE, and \emph{page-efficient} SSE.
The centerpiece of our approach is a novel connection between those two goals.
We introduce a map, called the \GLT, which takes as input a \emph{page-efficient} SSE scheme with certain special features, and outputs an SSE scheme with strong \emph{locality} properties.
We obtain several results.

\begin{itemize}
\item First, for page-efficient SSE, we build a \emph{dynamic} scheme with page efficiency $\bigO{\log \log N}$ and storage efficiency $\bigO{1}$, called $\pageSSE$. The main technical innovation behind $\pageSSE$ is a new weighted extension of the two-choice allocation process, of independent interest.
\item Second, we introduce the \GLT, and combine it with $\pageSSE$ to build a \emph{dynamic} SSE scheme with storage efficiency $\bigO{1}$, locality $\bigO{1}$, and read efficiency $\bigOtilde{\log\log N}$, under the condition that the longest list is of size $\bigO{N^{1-1/\log \log \lambda}}$. This matches, in every respect, the purely \emph{static} construction of Asharov et al. presented at STOC 2016: dynamism comes at no extra cost.
\item Finally, by applying the \GLT to a variant of the Tethys scheme by Bossuat et al. from Crypto 2021, we build an unconditional static SSE with storage efficiency $\bigO{1}$, locality $\bigO{1}$, and read efficiency $\bigO{\log^\varepsilon N}$, for an arbitrarily small constant $\varepsilon > 0$.
To our knowledge, this is the construction that comes closest to the lower bound presented by Cash and Tessaro at Eurocrypt 2014.
\end{itemize}


\end{abstract}


\section{Introduction}
\label{sec:intro}


{\bf Searchable Symmetric Encryption.}
In Searchable Symmetric Encryption (SSE), a client outsources the storage of a set of documents to an unstrusted server. The client wishes to retain the ability to search the documents, by issuing search queries to the server.
In the setting of \emph{dynamic} SSE, the client may also issue update queries, in order to modify the contents of the database, for instance by adding or removing entries.
The server must be able to correctly process all queries, while learning as little information as possible about the client's data and queries.
SSE is relevant in many cloud storage scenarios: for example, in cases such as outsourcing the storage of a sensitive database, or offering an encrypted messaging service, some form of search functionality may be highly desirable.

In theory, SSE is a special case of computation on encrypted data, and could be realized using generic solutions, such as Fully Homomorphic Encryption. In practice, such approaches incur a large performance penalty. Instead, SSE schemes typically aim for high-performance solutions, scalable to large real-world databases. Towards that end, SSE trades off security for efficiency. The server is allowed to learn some information about the client's data. For example, SSE schemes typically leak to the server the repetition of queries (\emph{search pattern}), and the identifiers of the documents that match a query (\emph{access pattern}). The security model of SSE is parametrized by a \emph{leakage function}, which specifies the nature of the information leaked to the server.

{\bf Locality.}
In the case of single-keyword SSE, search queries ask for all documents that contain a given keyword. To realize that functionality, the server maintains an (encrypted) reverse index, where each keyword is mapped to the list of the identifiers of documents that match the keyword. When the client wishes to search for the documents that match a given keyword, the client simply retrieves the corresponding list from the server.
A subtle issue, however, is how the lists should be stored and accessed by the server.

The naive approach of storing one list after the other is unsatisfactory: indeed, the position of a given list in memory becomes dependent on the lengths of other lists, thereby leaking information about those lists. A common approach to address that issue is to store each list element at a random location in memory. In that case, when retrieving a list, the server must visit as many random memory locations as the number of elements in the list. This is also undesirable, for a different reason: for virtually all modern storage media, accessing many random memory locations is much more expensive than visiting one continuous region.
Because SSE relies on fast symmetric cryptographic primitives, the cost of memory accesses becomes the performance bottleneck.
To capture that cost, \cite{EC:CasTes14} introduces the notion of \emph{locality}: in short, the locality of an SSE scheme is the number of discontinuous memory locations that the server must access to answer a query.

The two extreme solutions outlined above suggest a conflict between security and locality. At Eurocrypt 2014, Cash and Tessaro showed that this conflict is inherent~\cite{EC:CasTes14}: if a secure SSE scheme has constant storage efficiency (the size of the encrypted database is linear in the size of the plaintext database), and constant read efficiency (the amount of data read by the server to answer a search query is linear in the size of the plaintext answer), then it cannot have constant locality.

{\bf Local SSE constructions.}
Since then, many SSE schemes with constant locality have been proposed, typically at the cost of superconstant read efficiency.
At STOC 2016, \ash presented a scheme with $\bigO{1}$ storage efficiency, $\bigO{1}$ locality, and $\bigOtilde{\log N}$ read efficiency, where $N$ is the size of the database~\cite{STOC:ANSS16}.
At Crypto 2018, Demertzis et al. improved the read efficiency to $\bigO{\log^{2/3+\varepsilon} N}$~\cite{C:DemPapPap18}.
Several trade-offs with $\omega(1)$ storage efficiency were also proposed in \cite{demertzis2017fast}.
When the size of the longest list in the database is bounded, stronger results are known.
When such an upper bound is required, we will call the construction \emph{conditional}.
The first conditional SSE is due to \ash, and achieves $\bigOtilde{\log \log N}$ read efficiency, on the condition that the size of the longest list is $\bigO{N^{1-1/\log \log N}}$.
This was later improved to $\bigOtilde{\log \log \log N}$ read efficiency, with a stronger condition of $\bigO{N^{1-1/\log \log \log N}}$ on the size of the longest list.

Locality was introduced as a performance measure for memory accesses, assuming an implementation on Hard Disk Drives. In \cite{C:BBFMR21}, Bossuat et al. show that in the case of Solid State Drives (such as flash disks), locality is no longer the relevant target. Instead, performance is mainly determined by the number of memory pages accessed, regardless of whether they are contiguous. In that setting the right performance metric is \emph{page efficiency}. Page efficiency is defined as the number of pages read by the server to answer a query, divided by the number of pages needed to store the plaintext answer.
The main construction of \cite{C:BBFMR21} achieves $\bigO{1}$ storage efficiency and $\bigO{1}$ page efficiency, assuming a client-side memory of $\omega(\log \lambda)$ pages.

To this day, a common point among all existing constructions, both local and page-efficient, is that they are purely \emph{static}.
That may be because of the difficulty inherent in building local SSE, even in the static case (as evidenced, from the onset, by the impossibility result of Cash and Tessaro \cite{EC:CasTes14}). Nevertheless, many, if not most, applications of SSE require dynamism. This state of affairs significantly hinders the applicability of local and page-efficient SSE.

\subsection{Our Contributions}


In this article, we consider, for the first time, the problem of dynamic memory-efficient SSE, by which we mean that we target both dynamic \emph{page-efficient} SSE, and dynamic \emph{local} SSE.
The centerpiece of our approach is a novel connection between these two goals.
We introduce a map, called the \GLT, which takes as input a page-efficient SSE scheme with certain special features, and outputs an SSE scheme with strong locality properties.
Our strategy will be to first build page-efficient schemes, then apply the \GLT to obtain local schemes.
This approach turns out to be quite effective, and we present several results.


\begin{table}[!h]
    \caption{Page-efficient SSE schemes. $N$ denotes the total size of the database, $p$ is the number elements per page, $\varepsilon > 0$ is an arbitrarily small constant, and $\lambda$ is the security parameter.
    Page efficiency, storage efficiency, and client storage are defined in \Cref{sec:measures}.
    }
    
    \label{fig:comparePEF}
    \begin{center}
        \begin{tabular}{|l|r|r|r|r|r|}\hline
            Schemes                                   & Client st.                      & Page eff.           & Storage eff.      & Dynamism & Source\\\hline
            $\Pi_{\mathrm{pack}}, \Pi_{\text{2lev}}$ & $\bigo(1)$           & $\bigo(1)$         & $\bigo(p)$       & Static & \cite{NDSS:CJJJKR14}\\
            TCA                     & $\bigo(1)$          & $\bigOtilde{\log \log N}$ & $\bigo(1)$       & Static & \cite{STOC:ANSS16}\\
            Tethys                & $\bigo(p\log \lambda)$    & $3$                 & $3+\varepsilon$   & Static & \cite{C:BBFMR21}\\\hline
            $\pageSSE$                & $\bigO{1}$    & $\bigO{\log \log N}$                 & $\bigO{1}$   & Dynamic & \Cref{sub:sec_page_sse}\\
            \hline
        \end{tabular}
    \end{center}
\end{table}

\begin{table}[!h]
    \caption{SSE schemes with constant locality and storage efficiency. $N$ denotes the total size of the database, and $\varepsilon > 0$ is an arbitrarily small constant.
    Locality, read efficiency, and storage efficiency are defined in \Cref{sec:measures}.
    }
    
    \label{fig:compareLOC}
    \begin{center}
    \setlength\tabcolsep{0.33em}
        \begin{tabular}{|l|r|r|r|r|r|r|}\hline
            Schemes                                   & Locality           & Read eff. & St. eff.      & Max list size & Dynamism & Source\\\hline
            TCA                     & $\bigO{1}$ & $\bigOtilde{\log \log N}$ & $\bigo(1)$       & $\bigO{N^{1-1/\log \log N}}$ & Static & \cite{STOC:ANSS16}\\
            Leveled scheme                     & $\bigO{1}$ & $\bigOtilde{\log \log \log N}$ & $\bigo(1)$       & $\bigO{N^{1-1/\log \log \log N}}$ & Static & \cite{asharov2021tight}\\
            OCA                     & $\bigO{1}$ & $\bigOtilde{\log N}$ & $\bigo(1)$       & Unconditional & Static & \cite{STOC:ANSS16}\\
            DPP18                     & $\bigO{1}$ & $\bigOtilde{\log^{2/3+\varepsilon} N}$ & $\bigo(1)$       & Unconditional & Static & \cite{C:DemPapPap18}\\\hline
            $\local[\pageSSE]$                & $\bigO{1}$    & $\bigOtilde{\log \log N}$                 & $\bigO{1}$   & $\bigO{N^{1-1/\log \log N}}$ & Dynamic & \Cref{subsec:GLT}\\
            $\uselessSSE$                & $\bigO{1}$    & $\bigOtilde{\log^\varepsilon N}$                 & $\bigO{1}$   & Unconditional & Static & \Cref{sec:useless}\\
            \hline
        \end{tabular}
    \end{center}
\end{table}

\begin{itemize}
\item[--] {\bf Dynamic page-efficient SSE.}
We start by building a dynamic page-efficient SSE scheme, $\pageSSE$. $\pageSSE$ achieves storage efficiency $\bigO{1}$, and page efficiency $\bigO{\log \log N}$.
In line with prior work on memory-efficient SSE, the technical core of $\pageSSE$ is a new dynamic allocation scheme, $\layeredchoice$. $\layeredchoice$ is a weighted variant of the so-called ``2-choice'' algorithm, notorious in the resource allocation literature. (More details are provided in the technical overview.) As such, $\layeredchoice$ is of independent interest.
\item[--] {\bf \GLT.}
We introduce the \GLT. On input any page-efficient scheme $\peff$ with certain special features, called \emph{page-length-hiding} SSE, the \GLT outputs a local SSE scheme $\local{\peff}$. Roughly speaking, if $\peff$ has client storage $\bigO{1}$, storage efficiency $\bigO{1}$, and page efficiency $\bigO{P}$, then $\local{\peff}$ has storage efficiency $\bigO{1}$, and read efficiency $\bigO{P}$. Regarding locality, the key feature is that if $\peff$ has locality $\bigO{L}$ \emph{when querying lists of size at most one page}, then $\local{\peff}$ has locality $\bigO{L + \log \log N}$ \emph{when querying lists of any size}.
Thus, the $\local$ construction may be viewed as bootstrapping a scheme with weak locality properties into a scheme with much stronger locality properties.

The \GLT also highlights an interesting connection between the goals of page efficiency and locality. Originally, locality and page efficiency were introduced as distinct performance criterions, targeting the two most widespread storage media\textemdash respectively, Hard Disk Drives, and Solid State Drives. It was already observed in \cite{C:BBFMR21} that a scheme with locality $L$ and read efficiency $R$ must have page efficiency at most $R+2L$. In that sense, page efficiency is an ``easier'' goal. With the \GLT, suprisingly, we build a connection in the reverse direction: we use page-efficient schemes as building blocks to obtain local schemes. On a theoretical level, this shows a strong connection between the two goals. On a practical level, it provides a strategy to target both goals at once.

\item[--] {\bf Dynamic local SSE.}
By applying the \GLT to the $\pageSSE$ page-efficient scheme, we immediately obtain a dynamic SSE scheme $\local{\pageSSE}$, with storage efficiency $\bigO{1}$, locality $\bigO{1}$, and read efficiency $\bigO{\log \log N}$. The construction is conditional: it requires that the longest list is of size $\bigO{N^{1-1/\log \log N}}$. The asymptotic performance of $\local{\pageSSE}$ matches exactly the second \emph{static} construction from \cite{STOC:ANSS16}, including the condition on maximum list size: dynamism comes at no extra cost.
In particular, $\local{\pageSSE}$ matches the lower bound from \cite{asharov2021tight} for SSE schemes built using what \cite{asharov2021tight} refers to as ``allocation schemes''\textemdash  showing that the bound can be matched even in the dynamic setting.

\item[--] {\bf Unconditional local SSE in the static setting.}
The original 1-choice scheme from \cite{STOC:ANSS16} achieves $\bigO{1}$ storage efficiency, $\bigO{1}$ locality, and $\bigOtilde{\log N}$ read efficiency, unconditionally. The read efficiency was improved to $\bigO{\log^{2/3+\varepsilon} N}$ in \cite{C:DemPapPap18}, for any constant $\varepsilon > 0$. This was, until now, the only SSE construction to achieve sublogarithmic efficiency unconditionally.
By applying the \GLT to a variant of Tethys \cite{C:BBFMR21}, in combination with techniques inspired by \cite{C:DemPapPap18}, we obtain an unconditional static SSE scheme with storage efficiency $\bigO{1}$, locality $\bigO{1}$, and read efficiency $\bigO{\log^{\varepsilon} N}$, for any constant $\varepsilon > 0$.
To our knowledge, this is the construction that comes closest to the impossibility result of Cash and Tessaro, stating that $\bigO{1}$ locality, storage efficiency, and read efficiency simultaneously is impossible.
\end{itemize}

\paragraph{Remark on Forward Security.}
The SSE schemes built in this work have a standard ``minimal'' leakage profile during $\Search$: namely, searches leak the search pattern and access pattern.
For our dynamic schemes, $\Update$ operations leak an identifier of the list being updated, as well as, in some cases, the length of the list.
As a consequence, our dynamic schemes are not \emph{forward-secure}.
The underlying issue is that the goals of forward security and memory efficiency seem to be fundamentally at odds.
Indeed, locality asks that identifiers associated to the same keywords must be stored close to each other; while forward-privacy requires that the location where a new identifier is inserted should be independent of the keyword it is associated with.
That issue was already noted in \cite{CCS:Bost16}, who claims that ``for dynamic schemes, locality and forward-privacy are two irreconcilable notions''.
We refer the reader to  \cite{CCS:Bost16} for more discussion of the problem.
We leave further analysis of this issue for future work.

\section{Technical Overview}
\label{sec:tech}


This work contains several results, tied together by the \GLT.
As such, we believe it is beneficial to present them together within one paper.
This requires introducing a number of different allocation mechanisms.
We have endeavored to provide in this section a clear overview of those mechanisms.
Formal specifications, theorems, and proofs will be presented in subsequent sections.

It is helpful to fist recall a few well-studied allocation mechanisms. In what follows, ``with overwhelming probability'' is synonymous with ``except with negligible probability'' (in the usual cryptographic sense), whereas ``with high probability'' simply means with probability close to 1 in some sense, but not necessarily overwhelming.

\paragraph{One-choice allocation.}
In one-choice allocation, $n$ balls are thrown into $n$ bins. Each ball is inserted into a bin chosen independently and uniformly at random (by hashing an identifier of the ball). A standard analysis using Chernoff bounds shows that, at the outcome of the insertion process, the most loaded bin contains $\bigO{\log n}$ balls with high probability~\cite{johnson1977urn}. (And at most $\bigO{f(n)\log n}$ balls with overwhelming probability, for any $f = \omega(1)$.)

\paragraph{Two-choice allocation.}
Once again, $n$ balls are thrown into $n$ bins.
For each ball, two bins are chosen independently and uniformly at random (e.g. by hashing an identifier of the ball).
The ball is inserted into whichever of the two bins contains the fewest balls at the time of insertion.
A celebrated result by Azar et al. shows that, at the outcome of the insertion process, the most loaded bin contains $\bigO{\log \log n}$ balls with high probability~\cite{azar}.
(It was later shown that the result holds with overwhelming probability~\cite{richa2001power}.)

\paragraph{Cuckoo hashing.}
Cuckoo hashing is a classic hashing scheme introduced by Pagh and Rodler~\cite{cuckoo}. It has found many applications within cryptography: among others, oblivious algorithms (cf.~\cite{AC:CGLS17}, and the references therein), private set intersection~\cite{USENIX:PSSZ15}, and more recently, searchable encryption~\cite{CCS:PPYY19,C:BBFMR21}.
In cuckoo hashing, $n$ balls are inserted into $(2+\varepsilon)n$ \emph{cells}, where $\varepsilon > 0$ is an arbitrarily small constant.
Each cell can contain at most one ball. For each ball, two cells are chosen independently and uniformly at random (e.g. by hashing an identifier of the ball).
The ball is inserted into one of the two cells. If the cell was already occupied, the occupying ball is moved to its other possible destination cell, possibly creating a chain reaction.
Pagh and Rodler have shown that insertion terminates in expected $\bigO{\log n}$ time~\cite{cuckoo} (including the amortized cost of rebuilding the whole table with a new hash function in case of insertion failure).
In the end, similar to two-choice allocation, each ball is stored in one of two possible locations. Thanks to the more complex insertion algorithm, which allows moving already placed balls, the most loaded cell has (by definition) a load of 1, instead of $\bigO{\log \log n}$ for two-choice allocation.
To achieve a negligible probability of failure, cryptographic applications typically use cuckoo hashing with a stash~\cite{SIAM:KirMitWie10}.

\subsection{Layered 2-Choice Allocation}

Our first goal is to build a dynamic page-efficient scheme.
Let us summarize what this entails, starting with the static case.
As explained in the introduction, to realize single-keyword SSE, we want to store lists of arbitrary sizes on an untrusted server. Hiding the contents of the lists can be achieved in a straightforward way using symmetric encryption. The main challenge is how to store the lists in the server memory, in such a way that accessing one list does not reveal information about the lengths of other lists.

In the case of page-efficient schemes, this challenge may be summarized as follows.
We are given a set of lists, containing $N$ items in total. We are also given a page size $p$, which represents the number of items that can fit within a physical memory page.
The memory of the server is viewed as an array of pages. We want to store the lists in the server memory, with three goals in mind.
\begin{enumerate}
\item In order to store all lists, we use $S \lceil N/p\rceil$ pages of server memory in total, where $S$ is called the \emph{storage efficiency} of the allocation scheme. We want $S$ to be as small as possible.
\item Any list of length $\ell$ can be retrieved by visiting at most $P\lceil \ell/p \rceil$ pages in server memory, where $P$ is called the \emph{page efficiency} of the allocation scheme. We want $P$ to be as small as possible.
\item Finally, the pages visited by the server to retrieve a given list should not depend on the lengths of other lists.
\end{enumerate}
The first two goals are precisely the aim of bin packing algorithms. The third goal is a security goal: it stipulates that the pattern of memory accesses performed by the server should not leak certain information.
As such, the goal relates to oblivious or data-independent algorithms.
In \cite{C:BBFMR21}, a framework for realizing the three goals was formalized as \emph{Data-Independent Packing} (DIP).

To ease presentation, we will focus on the case where all lists are of size at most one page.
If a list is of length more than one page, the general idea is that it will be split into chunks of one page, plus one final chunk of size at most one page; each chunk will then be treated as a separate list by the allocation scheme. We assume from now on that lists are of length less than one page.

In a nutshell, the idea proposed by \cite{C:BBFMR21} to instantiate a DIP scheme is to use weighted variant of cuckoo hashing.
In more detail,
for each list, two pages are chosen uniformly at random, by hashing an identifier of the list. Each element of the list will then be stored in one of the two designated pages, or a stash. The stash is stored on the client side. In order to choose how each list is split between its three possible destinations (the two chosen pages, or the stash), \cite{C:BBFMR21} uses a maximum flow algorithm.  The details of this algorithm are not relevant for our purpose. The important point is that
when retrieving a list, the server accesses two uniformly random pages. Clearly, this reveals no information to the server about the lengths of other lists. The resulting algorithm, called Tethys, achieves storage efficiency $\bigO{1}$, page efficiency $\bigO{1}$, with client storage $\omega(\log \lambda)$ pages (used to store the stash).

In this paper, we wish to build a dynamic SSE.
For that puporse, the underlying allocation scheme needs to allow for a new \emph{update} operation. An update operation allows the client to add a new item to a list, increasing its length by one.
The security goal remains essentially the same as in the static case: the pages accessed by the algorithm in order to update a given list should not depend on the lengths of other lists.

Tethys is not a suitable basis for a dynamic scheme, because it does not allow for an efficient data-independent update procedure: when inserting an element into a cell during an update, the update procedure requires accessing other cells, with an access pattern that is intrinsically data-dependent.
Instead, a natural idea is to use a weighted variant of the two-choice allocation scheme.
With two-choice allocation, the access pattern made during an update is simple: only the two destination buckets associated to the list being updated need to be read.
The new item is then inserted into whichever of the two buckets currently contains less items.

Instantiating that approach would require a weighted variant of two-choice allocation, along the following lines: given a multiset of list sizes $\{\ell_i : 1 \leq i \leq k\}$ with $\ell_i \leq p$ and $\sum \ell_i = N$, at the outcome of a two-choice allocation process into $\bigO{N/p}$ buckets, the most loaded bucket contains $\bigO{p \log \log N}$ items with overwhelming probability.
However, a result of that form appears to be a long-standing open problem (some related partial results are discussed in~\cite{berenbrink2008weighted}).
The two-choice process with weighted items has been studied in the literature~\cite{talwar2007balanced,talwar2014balanced}, but to our knowledge, all existing results assume that the weight of the balls are sampled identically and independently from a sufficiently smooth distribution. Even disregarding constraints on the distribution, in our setting, we cannot even afford to assume that list lengths are drawn independently: in the SSE security model, lists are chosen and updated \emph{arbitrarily} by the adversary.

For our purpose, we require a \emph{distribution-free} statement: we only know a bound $p$ on the size of each list, and a bound $N$ on the total size of all lists.
We want an  $\bigO{p \log \log N}$ upper bound on the size of the most loaded bucket that holds for \emph{any} set of list sizes satisfying those constraints.
A result of that form is known for one-choice allocation processes \cite{berenbrink2008weighted} (with a $\bigO{p \log N}$ upper bound), but the same article shows that the same techniques cannot extend to the two-choice process.

To solve that problem, we introduce a \emph{layered} weigthed 2-choice allocation algorithm, $\layeredchoice$. $\layeredchoice$ has the same basic behavior as a (weighted) two-choice algorithm: for each ball, two bins are chosen uniformly at random as possible destinations. The only difference is how the bin where the ball is actually inserted is selected among the two destination bins. The most natural choice would be to store the ball in whichever bin currently has the least load, where the \emph{load} of a bin is a the sum of the weights of the balls it currently contains. Instead, we use a slightly more complex decision process. In a nutshell, we partition the possible weights of balls into $\bigO{\log \log \lambda}$ subintervals, and the decision process is performed independently for balls in each subinterval. For the first subinterval (holding the smallest weights), we use a weighted one-choice process, while for the other subintervals, we use an unweighted two-choice process.

The point of this construction is that its analysis reduces to the analysis of the weighted one-choice process, and the unweighted two-choice process, for which powerful analytical techniques are known.
We leverage those techniques to show that $\layeredchoice$ achieves the desired distribution-free guarantees on the load of the most loaded bin.
In practice, what this means is that we have an allocation algorithm that, for most intents and purposes, behaves like a weighted variant of two-choice allocation, and for which distribution-free guarantees can be obtained relatively painlessly. Multiple-choice allocation processes are ubiquitous in some areas of computer science\footnote{Such as load balancing, hashing, job allocation, or circuit routing. The reader is referred to~\cite{richa2001power} for a survey. To further illustrate the point, Azar, Broder, Karlin, Mitzenmacher and Upfal have recently received the 2020 ACM Paris Kenallakis Theory and Practice Award for the discovery and analysis of the two-choice process, and its extensive applications to practice~\cite{kanellakis}.}, making this a result of independent interest.

The $\pageSSE$ scheme is obtained by adding a layer of encryption and key management on top of $\layeredchoice$, using standard techniques from the SSE literature, although some care is required for updates.
We refer the reader to \Cref{sub:sec_page_sse} for more details.

\subsection{\GLT}

At Crypto 2018, \ash identified two main paradigms for building local SSE~\cite{C:AshSegSha18}. The first is the \emph{allocation} paradigm, which typically uses variants of multiple-choice allocation schemes, or cuckoo hashing. The second is the \emph{pad-and-split} approach. The main difficulty of memory-efficient SSE is to pack together lists of different sizes. The idea of the pad-and-split approach is to store lists separately according to their size, which circumvents the issue. The simplest way to realize this is to pad all lists length to the next power of 2. This yields $\log N$ possible values for list lengths. All lists of a given length can be stored together using, for instance, a standard hash table. Since we do not want to reveal the number of lists of each length, the hash table at each level needs to be dimensioned to be able to receive the entire database. As a result, a basic pad-and-split scheme has storage efficiency  $\bigO{\log N}$, but easily achieves $\bigO{1}$ locality and read efficiency.

For the \GLT, we introduce the notion of \emph{Overflowing SSE} (OSSE). An OSSE behaves like an SSE scheme in all aspects, except that, during its setup and during updates, it may refuse to store some list elements. Such elements are called \emph{overflowing}.
An OSSE is intended to be used as a subcomponent within an overaching SSE construction. The OSSE scheme is used to store part of the database, while overflowing elements are stored using a separate mechanism.
The notion of OSSE was not formalized before, but in hindsight, the use of OSSE may be viewed as implicit in several existing constructions \cite{C:DemPapPap18,C:AshSegSha18,C:BBFMR21}. We choose to introduce it explicitly here for ease of exposition.

We are now in a position to explain the \GLT.
The chief limitation of the pad-and-split approach is that it creates a $\log N$ overhead in storage. The high-level idea of the \GLT, then, is to use an OSSE to store all but a fraction $1/\log N$ of the database. Then a pad-and-split variant is used to store the $N/\log N$ overflowing elements. The intent is to benefit from the high efficiency of the pad-and-split approach, without having to pay for the $\log N$ storage overhead.

There is, however, a subtle but important issue with that approach.
A given list may be either entirely stored within the OSSE scheme, or only partially stored, or not stored at all. In the OSSE scheme that we will later use (as well as OSSEs that were implicit in prior work), those three situations should be indistinguishable to the server, or else security breaks down.
To address that issue, we proceed as follows.

Let us assume all lists have been padded to the next power of 2. For the pad-and-split part of the construction, we create $\log N$ SSE instances, one for each possible list size.
We call each of these instances a \emph{layer}.
If a list is of size $\ell$ its overflowing elements will be stored in the layer that handles lists of size $\ell$, regardless of how many elements did overflow from the OSSE for that list.

The only guarantee provided by the OSSE regardling overflowing elements is that their total number is $n = \bigO{N/\log N}$.
Thus, if we focus on the layer that handles lists of size $\ell$, the layer will receive at most $n$ elements.
These elements will be split into lists of size at most $\ell$ (corresponding to the set of overflowing elements, for each list of size $\ell$ in the original database).
To achieve storage efficiency $\bigO{S}$ overall, we want the layer to store those lists using $\bigO{Sn}$ storage. To achieve read efficiency $R$, the layer should also be able to retrieve a given list by visiting at most $R\ell$ memory locations.
This is where everything comes together: an SSE scheme satisfying those conditions is precisely a page-efficient SSE scheme with page size $\ell$, storage efficiency $S$, and page efficiency $R$.

The page-efficient scheme used for each layer is also required satisfy a few extra properties: first, when searching for a list of size at most one page, the length of the list should not be leaked.
We call this property \emph{page-length-hiding}. (We avoid the term \emph{length-hiding} to avoid confusion with volume-hiding SSE, which fully hides lengths.)
All existing page-efficient constructions have that property. Second, we require the page-efficient scheme to have $\bigO{1}$ client storage. All constructions in this article satisfy that property, but the construction from \cite{C:BBFMR21} does not. Finally, we require the scheme to have locality $\bigO{1}$ when fetching a single page. All existing page-efficient constructions have this property. (The last two properties could be relaxed, at the cost of more complex formulas and statements.)
We call an SSE scheme satisfying those three properties \emph{suitable}.

Putting everything together, the \GLT takes as input a suitable \emph{page-efficient} scheme, with storage efficiency $S$ and page efficiency $P$. It outputs a \emph{local} scheme with storage efficiency $S + S'$, read efficiency $P + R'$, and locality $L'$, where $S'$, $R'$, and $L'$ are the storage efficiency, read efficiency, and locality of the underlying OSSE.
It remains to explain how to build a local OSSE scheme with $\bigO{N/\log N}$ overflowing items, discussed next.

\subsection{$\clippedOSSE$: an OSSE scheme with $\bigO{N/\log N}$ Overflowing Items}

At STOC 2016, \ash introduced so-called ``2-dimensional'' variants of one-choice and two-choice allocation, for the purpose of building local SSE. The one-choice variant works as follows. Consider an SSE database with $N$ elements. Allocate $m = \bigOtilde{N/\log N}$ buckets, initially empty. For each list of length $\ell$ in the database, choose one bucket uniformly at random. The first element of the list is inserted into that bucket. The second element of the list is inserted into the next bucket (assuming a fixed order of buckets, which wraps around when reaching the last bucket), the third one into the bucket after that, and so on, until all list elements have been inserted. Thus, assuming $\ell \leq m$, all list elements have been placed into $\ell$ consecutive buckets, one element in each. An analysis very similar to the usual analysis of the one-choice process shows that with overwhelming probability, the most loaded bucket receives at most $\tau = \bigOtilde{\log N}$ elements. To build a static SSE scheme from this allocation scheme, each bucket is padded to the maximal size $\tau$ and encrypted. Search queries proceed in the natural way.

Such a scheme yields storage efficiency $\bigO{1}$, locality \bigO{1} (since retrieving a list amounts to reading consecutive buckets), and read efficiency $\bigOtilde{\log N}$ (since retrieving a list of length $\ell$ requires reading $\ell$ buckets, each of size $\tau = \bigOtilde{\log N}$).
To build $\clippedOSSE$, we start from the same premise, but ``clip'' buckets at the threshold $\tau = \bigOtilde{\log \log N}$. That is, each bucket can only receive up to $\tau$ elements. Elements that cannot fit are overflowing.

In the standard one-choice process, where $n$ balls are thrown i.i.d. into $n$ bins, it is not difficult to show that clipping bins at height $\tau = \bigO{\log \log n}$ results in at most $\bigO{n/\log n}$ overflowing elements with overwhelming probability. In fact, by adjusting the multiplicative constant in the choice of $\tau$, the number of overflowing elements can be made $\bigO{n/\log^d n}$ for any given constant $d$. We show that a result of that form still holds for (a close variant of) the 2-dimensional one-choice process outlined earlier. The result is conditional: it requires that the maximum list size is $\bigO{N/\polylog\: N}$. (A condition of that form is necessary, insofar as the result fails when the maximum list size gets close to $N/\log N$.) The proof of the corresponding theorem is the most technically challenging part of this work, and relies on the combination of a convexity argument with a stochastic dominance argument. An overview of the proof is given in \ref{sec:clippedOSSE}, so we omit more discussion here.

In the end, $\clippedOSSE$ achieves storage efficiency $\bigO{1}$, locality $\bigO{1}$, and read efficiency $\bigO{\log \log N}$, with $\bigO{N/\log^d N}$ overflowing elements (for any fixed constant $d$ of our choice),
under the condition that the maximum list size is $\bigO{N/\polylog\: N}$.
All applications of the \GLT in this article use $\clippedOSSE$ as the underlying OSSE.
(That is why we write $\local[\peff]$ for the \GLT applied to the page-efficient scheme $\peff$, and do not put the underlying OSSE as an explicit parameter.)

\subsection{Dynamic Local SSE with $\bigO{\log \log N}$ Overhead}

By using the \GLT with $\clippedOSSE$ as the underlying OSSE, and $\pageSSE$ as the page-efficient scheme, we obtain $\local[\pageSSE]$. The $\local[\pageSSE]$ scheme has storage efficiency $\bigO{1}$, locality $\bigO{1}$, and read efficiency $\bigOtilde{\log \log N}$. This result follows from the main theorem regarding the \GLT, and does not require any new analysis.

$\local[\pageSSE]$ is a conditional scheme: it requires that the longest list is of length $\bigO{N^{1-1/\log \log \secpar}}$. The reason is subtle. $\clippedOSSE$ by itself has a condition that the longest list is $\bigO{N/\polylog\: N}$, which is less demanding. The reason for the condition comes down to the fact that $\pageSSE$ only achieves a negligible probability of failure as long as the number of pages in the scheme is at least $\Omega(\lambda^{1/\log \log \lambda})$. More generally, the same holds for the number of bins in two-choice allocation processes in general, even the standard, unweighted process. The condition is optimal: \cite{asharov2021tight} shows that any sublogarithmic ``allocation-based'' scheme must be conditional, and gives a bound on the condition. $\local[\peff]$ matches that bound.

\subsection{Unconditional Static Local SSE with $\bigO{\log^\varepsilon N}$ Overhead}

The (static) Tethys scheme from \cite{C:BBFMR21} achieves storage efficiency $\bigO{1}$ and page efficiency $\bigO{1}$ simultaneously. It is also page-length-hiding. Since we have the \GLT at our disposal, it is tempting to apply it to Tethys. There is, however, one obstacle: Tethys uses $\omega(p \log \lambda)$ client memory, in order to store a stash on the client side. For the \GLT, we need $\bigO{1}$ client memory. To reduce the client memory of Tethys, a simple idea is to store the stash on the server side. Naively, reading the stash for every search would increase the page efficiency to $\omega(\log \lambda)$. To avoid this, we store the stash within an ORAM.

For that purpose, we need an ORAM with a failure probability of zero: indeed, since we may store as few as $\log \lambda$ elements in the ORAM, a correctness guarantee of the form $\negl[n]$ where $n = \log \lambda$ is the number items in the ORAM fails to be sufficient (it is not $\negl$). We also need the ORAM to have $\bigO{1}$ locality. An ORAM with these characteristics was devised in \cite{C:DemPapPap18}, motivated by the same problem. The ORAM from \cite{C:DemPapPap18} achieves read efficiency $\bigO{n^{1/3 + \varepsilon}}$, for any arbitrary constant $\varepsilon > 0$. It was already conjectured in \cite{C:DemPapPap18} that it could be improved to $\bigO{n^{\varepsilon}}$. We build that variant explicitly, and name it $\localORAM$. Roughly speaking, $\localORAM$ is a variant of the Goldreich-Ostrovsky hierarchical ORAM, with a constant number of levels. 

By putting the stash of Tethys within $\localORAM$ on the server side, we naturally obtain a page-efficient SSE scheme $\tethysoram$, with $\bigO{\log^\varepsilon \lambda}$ read efficiency, suitable for use within the \GLT. This yields a static local SSE for lists of size at most $N/\polylog\: N$. To handle larger lists, borrowing some ideas from \cite{C:DemPapPap18}, we group lists by size, and use again $\tethysoram$ to store them.
In the end, we obtain an unconditional SSE with $\bigO{1}$ store efficiency, $\bigO{1}$ locality, and $\bigO{\log^\varepsilon \lambda}$ read efficiency.

Comparing with the $\bigO{\log^{2/3+\varepsilon} \lambda}$ construction from \cite{C:DemPapPap18}, we note that the bottleneck of their construction comes from the allocation schemes the authors use for what they call ``small'' and ``medium'' lists. This is precisely the range where we use $\local[\tethysoram]$. Our construction essentially removes that bottleneck, so that the $\bigO{\log^\varepsilon \lambda}$ read efficiency bottlneck now comes entirely from the ORAM component.
\section{Preliminaries}
\label{sec:prelim}
Let $\secpar \in \N$ be the security parameter. For a probability distribution $X$, we denote by $x \gets X$ the process of sampling a value $x$ from the distribution.
Further, we say that $x$ is 
We denote by $[a,b]_\R$ the interval $\{x \in \R \mid a \leq x \leq b\}$ and extend this naturally to intervals of the form $[a,b)_\R,(a,b]_\R,(a,b)_\R$.

\subsection{Symmetric Searchable Encryption}
A database $\DB = \{w_i, (\ind_1, \dots, \ind_{\ell_i})\}_{i=1}^W$ is a set of keyword-identifier pairs with $W$ keywords. 
We assume that each keyword $w_i$ is represented by a machine word of $\bigO{\secpar}$ bits.
We write $\DB(w_i) = (\ind_1, \dots, \ind_{\ell_i})$ for the list of identifiers matching $w_i$.
Also, we set $N = \sum_{i=1}^W \ell_i$. 
throughout the article, we define $p$ as the page size and we treat $p$ as a variable, independent of the size of the database $N$.

A dynamic searchable symmetric encryption scheme $\Sigma$ is a $4$-tuple of PPT algorithms $(\KeyGen,\allowbreak\Setup,\allowbreak\Search,\allowbreak\Update)$ such that
\begin{itemize}
    \item $\Sigma.\KeyGen(1^\secpar)$: Takes as input the security parameter $\secpar$ and outputs client secret key $\K$. 
    \item $\Sigma.\Setup(\K,N,\DB)$: Takes as input the client secret key $\K$, an upper bound on the database size $N$ and a database $\DB$. Outputs encrypted database $\EDB$ and client state $\st$. 
    \item $\Sigma.\Search(\K,w,\st;\EDB)$: The client receives as input the secret key $\K$, keyword $w$ and state $\st$. 
    The server receives as input the encrypted database $\EDB$. 
    Outputs some data $d$ and updated state $\st'$ for the client. 
    Outputs updated encrypted database $\EDB'$ for the server.
    \item $\Sigma.\Update(\K,(w,L'),\op,\st;\EDB)$: The client receives as input the secret key $\K$, a pair $(w,L)$ of keyword $w$ and list $L'$ of identifiers, an operation $\op\in\{\del,\add\}$ and state $\st$.
    The server receives as input the encrypted database $\EDB'$. 
    Outputs updated state $\st'$ for the client. 
    Outputs updated encrypted database $\EDB'$ for the server.
\end{itemize}
In the following, we omit the state $\st$ and assume that it is implicitly stored and updated by the client.
We say that $\Sigma$ is static, if it does not provide an $\Update$ algorithm. 
Further, we assume that the keyword $w$ is preprocessed via a \PRF by the client, whenever the client sends $w$ to the server in either $\Search$ or $\Update$.
This ensures that the server never has access to $w$ in plaintext and unqueried keywords are distributed uniformly random in the view of the server.

Intuitively, the client uses $\Sigma.\Setup$ to encrypt and outsource a database $\DB$ to the server.
Then, the client can search keywords $w$ using $\Sigma.\Search$ and receives matching identifieres $\DB(w)$ from the server.
The list of matching identifiers can also be updated using $\Sigma.\Update$, provided that the size of the database stays below $N$.

\subsubsection{Security.}
We now define correctness and semantic security of SSE. 
Intuitively, correctness guarantees that a search always retrieves all matching identifiers and semantic security guarantees that the server only learns limited information (quantified by a leakage function) from the client. 

\begin{definition}[Correctness]
A $\SSE$ scheme $\Sigma$ is correct if for all databases $\DB$ and $N\in\N$, keys $\K \gets \Sigma.\KeyGen(1^\secpar)$, $\EDB \gets \Sigma.\Setup(\K,\DB)$ and sequences of search, add or delete queries $S$, the search protocol returns the correct result for all queries of the sequence if the size of the database remains at most $N$.
\end{definition}

We use the standard semantic security notion for $\SSE$ (see \cite{CCS:CGKO06}). Security is parameterized by a leakage function $\L=(\L_\Stp,\L_\Srch,\L_\Updt)$, composed of the setup leakage $\L_\Stp$, the search leakage $\L_\Srch$, and the update leakage $\L_\Updt$.
We define two games, $\SSEReal$ and $\SSEIdeal$. 
First, the adversary chooses a database $\DB$. 
In $\SSEReal$, the encrypted database $\EDB$ is generated by $\Setup(\K,N,\DB)$, whereas in $\SSEIdeal$ the encrypted database is simulated by a (stateful) simulator $\Sim$ on input $\L_\Stp(\DB,N)$.
After receiving $\EDB$, the adversary issues search and update queries.
All queries are answered honestly in $\SSEReal$. 
In $\SSEIdeal$, the search queries on keyword $w$ are simulated by $\Sim$ on input $\L_\Srch(w)$ and update queries for operation $\op$, keyword $w$ and identifier list $L'$ are simulated by $\Sim$ on input $\L_\Updt(\op,w,L')$.
Finally, the adversary outputs a bit $b$.

We write $\SSEReal^\adaptive$ and $\SSEIdeal^\adaptive$ if the queries of the adversary were chosen adaptively, \ie dependant on previous queries.
Similarly, we write $\SSEReal^\selective$ and $\SSEIdeal^\selective$ if the queries are chosen selectively by the adversary, \ie sent initally in conjunction with the database before receiving $\EDB$.

\begin{definition}[Semantic Security]
Let $\Sigma$ be a $\SSE$ scheme and $\L = (\L_\Stp,\allowbreak\L_\Srch,\L_\Updt)$ a leakage function. Scheme $\Sigma$ is $\L$-adaptively secure if for all PPT adversaries $\adv$, there exists a PPT simulator $\Sim$ such that 
\[
    \abs{\Pr[\SSEReal^\adaptive_{\Sigma,\adv}(\secpar) = 1] - \Pr[\SSEIdeal^\adaptive_{\Sigma,\Sim,\L,\adv}(\secpar) = 1]} = \negl.
\]
Similarly, scheme $\Sigma$ is $\L$-selectively secure if for all PPT adversaries $\adv$, there exists a PPT simulator $\Sim$ such that 
\[
    \abs{\Pr[\SSEReal^\selective_{\Sigma,\adv}(\secpar) = 1] - \Pr[\SSEIdeal^\selective_{\Sigma,\Sim,\L,\adv}(\secpar) = 1]} = \negl.
\]
\end{definition}
Intuitively, semantic security guarantees that the interaction between client and server reveals no information to the server, except the leakage of the given query. 
The schemes from this article have common leakage patterns.
We use the standard notions of query pattern $\qpat$ and history $\Hist$ from \cite{CCS:Bost16} to formalize this leakage\footnote{\cite{CCS:Bost16} defines the search pattern and query pattern separately. We include the search pattern in the query pattern, since search queries can be linked to update queries on the same keyword without forward security, and vice versa.}: 
(1) The query pattern $\qpat(w)$ for a keyword $w$ are the indices of previous search or update queries for keyword $w$.
(3) The history $\Hist(w)$ is comprised of the list of identifiers matching keyword $w$ that were inserted during setup and the history of updates on keyword $w$, that is each deleted and inserted identifier. 
We can retreive the number $\ell_i$ of inserted identifiers and the number $d_i$ of deleted identifiers from $\Hist(w)$ for each keyword.

We define two leakage patterns we use throughout the article.
(1) We define page length hiding leakage $\LH$. We set $\LH = (\L_\Stp\lh, \L_\Srch\lh, \L_\Updt\lh)$, where the setup leakage is $\L_\Stp\lh(\DB, N) = N$ is the maximal size $N$ of the database, the search leakage $\L_\Srch\lh(w) = (\qpat, \ceil{\ell_i/p}, \ceil{d_i/p})$ is the query pattern and the number of pages required to store the inserted and deleted items, and the update leakage $\L_\Updt\lh(\op,w,L') = (\op,\qpat,\ceil{(\ell_i+\abs{L'})/p},\allowbreak\ceil{(d_i+\abs{L'})/p},\allowbreak\ceil{\ell_i/p},\ceil{d_i/p})$ is the operation, the query pattern and the number of pages required to store the inserted and deleted items (before and after the update)\footnote{Note that we allow for inserting more than one identifier per keyword in a single update operation in this work. Thus, the server will also learn (limited) information about the number of added or deleted identifiers $\abs{L'}$.}. 
(2) Similarly, we define length reveiling leakage $\LR$. We set $\LR = (\L_\Stp\lr, \allowbreak\L_\Srch\lr, \L_\Updt\lr)$ with $\L_\Stp\lr(\DB, N) = N$, $\L_\Srch\lr(w) = (\qpat, \abs{L'}, \ell_i, d_i)$ and lastly $\L_\Updt\lr(\op,w,L') = (\op,\qpat,\abs{L'},\ell_i,d_i)$. 

We will use $\LH$ and $\LR$ for both dynamic and static schemes. 
When we say that a static scheme is $\L$-semantically secure, for $\L\in\{\LH,\LR\}$, we simply ignore the update leakage.

\subsubsection{Efficiency Measures.}
\label{sec:measures}

We recall the notions of locality, storage efficiency and read efficiency~\cite{EC:CasTes14}, and page efficiency~\cite{C:BBFMR21}. Further, we extend them dynamic schemes in a natural manner.
In the following definitions, we set $\K \gets \KeyGen(1^{\secpar})$ and $\EDB \gets \Setup(\K,N,\DB)$ given database $\DB$ and upper bound $N$ on the number of document identifiers.
Also, $S = (\op_i,\opin_i)_{i=1}^s$ is a sequence of search and update queries, where $\op_i\in\{\add,\del,\bot\}$ is a operation and $\opin_i = (\op_i,w_i,L_i', \st_i,\EDB_i)$ its input. Here, $w_i$ is a keyword and $L'_i$ is a list of identifiers, and after executing all previous operations $\op_j$ for $j\leq i$, $\st_i$ is the client state and $\EDB_i$ the encrypted database. 
We denote by $\DB_i$ the database after $i$ operations.
We assume that the total number of identifiers never exceeds $N$. 
(If $\op_i = \bot$, the query is a search query and $L'_i$ is empty.)
We start with the definition of the \emph{read pattern}.

\begin{definition}[Read Pattern]
Regard server-side storage as an array of memory locations, containing the encrypted database $\EDB$.
When processing search query $\Search(\K,w_i,\st_i;\EDB_i)$ or update query $\Update(\K,(w_i,L'_i),\op_i,\st_i;\allowbreak\EDB_i)$, the server accesses memory locations $m_1,\dots, m_h$. 
We call these locations the \emph{read pattern} and denote it with $\RdPat(\op_i,\opin_i)$.
\end{definition}

\begin{definition}[Locality]
An SSE scheme has locality $L$ if for any $\lambda$, $\DB$, $N$, sequence $S$, and any $i$, $\RdPat(\op_i,\opin_i)$ consists of at most $L$ disjoint intervals.
\end{definition}

\begin{definition}[Read Efficiency]
An SSE scheme has read efficiency $R$ if for any $\lambda$, $\DB$, $N$, sequence $S$, and any $i$, $|\RdPat(\op_i,\opin_i)| \leq R \cdot P$,
where $P$ is the number of memory locations needed to store all (added and deleted) document indices matching keyword $w_i$ in plaintext (by concatenating indices).
\end{definition}

\begin{definition}[Storage Efficiency]
An SSE scheme has storage efficiency $E$ if for any $\lambda$, $\DB$, $N$, sequence $S$, and any $i$, $|\EDB_i| \leq E \cdot |DB_i|$. 
\end{definition}

Similarly, we now define page efficiency. This efficiency measure targets the storage medium SSD. 
\begin{definition}[Page Pattern]
Regard server-side storage as an array of \emph{pages}, containing the encrypted database $\EDB$.
When processing search query $\Search(\K,w_i,\st_i;\EDB_i)$ or update query $\Update(\K,(w_i,L'_i),\op_i,\st_i;\EDB_i)$, the read pattern $\RdPat(\op_i,\opin_i)$ induces a number of page accesses $p_1,\dots,p_{h'}$.
We call these pages the \emph{page pattern}, denoted by $\PgPat(\op_i,\opin_i)$.
\end{definition}

\begin{definition}[Page Cost]
An SSE scheme has page cost $aX + b$, where $a$, $b$ are real numbers, and $X$ is a fixed symbol,
if for any $\lambda$, $\DB$, $N$, sequence $S$, and any $i$, $|\PgPat(\op_i,\opin_i)| \leq a X + b$, where $X$ is the number of pages needed to store documents indices matching keyword $w_i$ in plaintext.
\end{definition}

\begin{definition}[Page Efficiency]
An SSE scheme has page efficiency $P$ if for any $\lambda$, $\DB$, $N$, sequence $S$, and any $i$, $|\PgPat(\tau,\EDB)| \leq P \cdot X$, where $X$ is the number of pages needed to store
documents indices matching keyword $w_i$ in plaintext.
\end{definition}

\section{Layered Two-Choice Allocation}
\label{sec:alloc}
In this section, we describe our algorithm $\layeredchoice$ that allows to allocate $n$ weighted balls into $m$ bins, where each ball $b_i$ has weight $w_i \in [0,1]_\R$.
First, let $1 \leq \delta(\secpar) \leq \log(\secpar)$ be a function.
We denote by $w = \sum_{i=1}^n w_i$ the sum of all weights and set $m = w/(\delta(\secpar)\log\log w)$.
We will later choose $\delta(\secpar) = o(\log\log \secpar)$ such that allocation has negligible failure probability later. 
In the overview, we set $\delta(\secpar) = 1$ and assume that $m = \Omega(\secpar)$ for simplicity (which suffices for negligible failure probability).

\subsubsection{Overview of $\layeredchoice$.}
\label{subsub:overviewl2c}
$\layeredchoice$ is based on both unweighted one-choice allocation (\onechoice) and unweighted two-choice allocation (\twochoice).
On a high level, we split the set of possible weights $[0, 1]_\R$ into $\log \log m$ subintervals
\[
[0,1/\log m]_\R, \allowbreak (1/\log m,2/\log m]_\R, \allowbreak \dots, \allowbreak (2^{\log\log m -1}/\log m, 1]_\R.
\]
In words, the first interval is size $1/ \log m$ and the boundaries between intervals grow by a factor 2 every time.
We will allocate balls with weights in a given subinterval independently from the others.

Balls in the first subinterval have weights $w_i \leq \log m$ and are thus small enough to apply $\onechoice$. 
Intuitively, this suffices because one-choice performs worst for uniform weights of maximal size $1/\log m$. In that case, there are at most $n' = w\log m$ balls and we expect a bin to contain $n'/m = \log m \cdot \log \log w$ balls of uniform weight, since $m = w/(\log \log w)$. 
As each ball has weight $1/\log m$, the expected load per bin is $\log \log w$. This translates to a $\bigO{\log \log w}$ bound with overwhelming probability after applying a Chernoff’s bound.

For the other intervals, applying unweighted and independent $\twochoice$ per interval suffices, as the weights of balls differ at most by a factor $2$ and there are only $\log \log m$ intervals. 
More concretely, let $n_i$ be the number of balls in the $i$-th subinterval $A_i = (2^{i-1}/\log m, 2^i/\log m]_\R$ for $i \in \Int{\log\log m}$. 
Balls with weights in subinterval $A_i$ fill the bins with at most $\bigO{n_i/m + \log\log m}$ balls, independent of other subintervals. Note that we are working with small weights, and thus potentially have $\omega(m)$ balls. Thus, we need to extend existing $\twochoice$ results to negligible failure probability in $m$ for the heavily-loaded case (cf. \cref{lemma:twochoice}).
As there are only $\log\log m$ subintervals and balls in interval $A_i$ have weight at most $2^i/\log m$, we can just sum the load of each subinterval and receive a bound 
\[ 
    \sum_{i=1}^{\log\log m} \frac{2^i}{\log m}\bigO{n_i/m+\log\log m} = \bigO{w/m + \log\log m}.
\]
In total, we have $\bigO{w/m + \log\log m} = \bigO{\log\log m}$ bounds for the first and the remaining intervals. Together, this shows that all bins have load at most $\bigO{\log\log m}$ after allocating all $n$ items.
This matches the bound of standard $\twochoice$ with unweighted balls if $m = \Omega(\secpar)$. 
For our SSE application, we want to allow for negligible failure probability with the least number of bins possible.
We can set $\delta(\secpar) = \log\log\log(\secpar)$ and obtain a bin size of $\bigOtilde{\log \log w}$ with overwhelming probability, if $m = \frac{w}{\delta(\secpar)\log\log w}$. The analysis is identical in this case.

\para{Handling Updates}
The described variant of $\layeredchoice$ is static. 
That is, we have not shown a bound on the load of the most loaded bin if we add balls or update the weight of balls.
Fortunately, inserts of new balls are trivially covered by the analysis sketched above, if $m$ was chosen large enough initially in order to compensate for the added weight.
Thus, we assume there is some upper bound $\maxweight$ on the total weights of added balls which is used to initially set up the bins.
We can also update weights if proceed with care. 

For this, let $b_i$ be some ball with weight $\oldweight$. We want to update its weight to $\newweight > \oldweight$.
If $\oldweight$ and $\newweight$ reside in the subinterval, we can directly update the weight of $b_i$, as $\layeredchoice$ ignores the concrete weight of balls inside a given subinterval for the choice of its bins.
Indeed, in the first interval, the bin in which $b_i$ is inserted is determined by a single random choice and for the remaining subintervals, the $\twochoice$ process only considers the number of balls inside the same subinterval, ignoring concrete weights.

When $\newweight$ is larger than the bounds of the current subinterval, we need to make sure that the ball is inserted into the correct bin of its two choices.
For this, the ball $b_i$ is inserted into the bin with the lowest number of balls with weights inside the new subinterval. 
Even though the bin of $b_i$ might change in this process, we still need to consider $b_i$ as a ball of weight $\oldweight$ in the old bin for subsequent ball insertions in the old subinterval.
Thus, we mark the ball as \emph{\leftover} ball but do not remove it from its old bin. 
That is, we consider it as ball of weight $\oldweight$ for the $\twochoice$ process but assume it is not identified by $b_i$ anymore.
As there are only $\log\log m$ different subintervals, the \leftover balls only have a constant overhead.
The full algorithm $\layeredchoice$ is given in \Cref{algo:layeredchoice}. 
We parameterize it by a hash function $\Hash$ mapping uniformly into $\Int{m}^2$. The random bin choices of a ball $b_i$ are given by $\alpha_1, \alpha_2 \gets \Hash(b_i)$.

\subsubsection{Load Analysis of $\layeredchoice$.}
Let either $\delta(\secpar) = 1$ or $\delta(\secpar) = \log\log\log \secpar$ and $m$ sufficiently large such that $m^{-\Omega(\delta(\secpar)\log\log w)} = \negl$. 
(Note that this is the probability that allocation of $\onechoice$ and $\twochoice$ fails.)

We need to show that after setup and during a (selective) sequence of operations, the most loaded bin has a load of at most $\bigO{\delta(\secpar)\log \log \maxweight}$, where $\maxweight$ is an upper bound on the total weight of the inserted balls. We sketch the proof here and refer to \Cref{sec:alloc_proof} for further details.
First, we modify the sequence $S$ such that we can reduce the analysis to only (sufficiently independent) $\layeredchoice.\InsertBall$ operations, while only increasing the final bin load by a constant factor.
This is constant factor of the load is due to the additional weight of \leftover balls. 
Then, we analyse the load of the most loaded bin for the each subinterval independently.
This boils down to an analysis of a $\onechoice$ process in the first subinterval and a $\twochoice$ process in the remaining subintervals as in the overview of $\layeredchoice$ (see \Cref{subsub:overviewl2c}). 
Summing up the independent bounds yields the desired result.

\begin{theorem}
\label{thm:layeredchoice}
Let either $\delta(\secpar) = 1$ or $\delta(\secpar) = \log\log\log \secpar$.
Let $\maxweight = \poly$ and $m =\maxweight/(\delta(\secpar)\log\log\maxweight)$. 
We require that $m = \Omega(\secpar^{\frac{1}{\log\log \secpar}})$ if $\delta(\secpar) = \log\log\log \secpar$ or $m = \Omega(\secpar)$ otherwise.
Let $\{(b_i, w_i)_{i=1}^n\}$ be balls with (pair-wise unique) identifier $b_i$ and weight $w_i \in [0,1]$.
Further, let $S = (\op_i, \opin_i)_{i=n+1}^{s+n}$ be a sequence of $s$ insert or update operations $\op_i \in \{\layeredchoice.\InsertBall,\layeredchoice.\UpdateBall\}$ with input $\opin_i = (b_i, w_i, B_{\alpha_{i,1}},B_{\alpha_{i,2}})$ for inserts and $\opin_i = (b_i, o_i, w_i, B_{\alpha_{i,1}},B_{\alpha_{i,2}})$ for updates. Here, $b_i$ denotes the identifier of a ball with weight $w_i$ and old weight $o_i \leq w_i$ before the execution of $\op_i$. 
Also, the bins are chosen via $\alpha_{i,1}, \alpha_{i,2} \gets \Hash(b_i)$.

Execute $(B_i)_{i=1}^m \gets \layeredchoice.\Setup(\{(b_i, w_i)_{i=1}^n\})$ and the operations $\op_i(\opin_i)$ for all $i\in[n+1,n+s]$.
We require that $\sum_{i=1}^{n+s} w_i - o_i \leq \maxweight$, \ie the total weight after all operations is at most $\maxweight$.

Then it holds that throughout the process, the most loaded bin of $B_1,\dots,B_m$ has at most load $\bigO{\delta(\secpar)\log \log \maxweight}$ except with negligible probability, if $\Hash$ is modeled as a random oracle.
\end{theorem}

\begin{algorithm}[h]
\begin{flushleft}
\caption{Layered 2-Choice Allocation (\layeredchoice)}
\label{algo:layeredchoice}
\underline{$\layeredchoice.\Setup(\{(b_i, w_i)_{i=1}^n\}, \maxweight)$}
\begin{algorithmic}[1]
\State Receive $n$ balls $b_i$ with weight $w_i$, and maximal total weight $\maxweight$
\State Initilize $m = \ceil{\maxweight/(\delta(\secpar)\log\log \maxweight)}$ empty bins $B_1,\dots,B_m$
\ForAll{$i\in\Int{n}$}
\State Set $\alpha_1, \alpha_2 \gets \Hash(b_i)$
\State $\InsertBall(b_i, w_i, B_{\alpha_1},B_{\alpha_2})$
\EndFor
\State Return $B_1,\dots,B_m$
\end{algorithmic}
\underline{$\layeredchoice.\InsertBall(\newball, \newweight, B_{\alpha_1},B_{\alpha_2})$}
\begin{algorithmic}[1]
\State Receive bins $B_{\alpha_1},B_{\alpha_2}$ and ball $\newball$ with weight $\newweight$
\State Assert that $\alpha_1, \alpha_2$ are the choices given by $\Hash(\newball)$
\State Split the set of possible weights $[0,1]_\R$ into $\log \log m$ sub-intervals
\[
[0,1/\log m]_\R, \allowbreak (1/\log m,2/\log m]_\R, \allowbreak \dots, \allowbreak (2^{\log\log m -1}/\log m, 1]_\R
\]
\State Choose $k\in\N$ minimal such that $\newweight \leq 2^k/\log m$
\If{k = 1}
\State Set $\alpha \gets \alpha_1$
\Else
\State \multiline{Let $B_\alpha$ be the bin with the least number of balls of weight in $\left(\frac{2^{k-1}}{\log m},\frac{2^{k}}{\log m}\right]_\R$ among $B_{\alpha_1}$ and $B_{\alpha_2}$}
\EndIf
\State Insert ball $\newball$ into bin $B_\alpha$
\end{algorithmic}
\underline{$\layeredchoice.\UpdateBall(\oldball, \oldweight, \newweight, B_{\alpha_1},B_{\alpha_2})$}
\begin{algorithmic}[1]
\State \multiline{Receive bins $B_{\alpha_1},B_{\alpha_2}$ that contain ball $\oldball$ with weight $\oldball$ and the new weight $\newweight \geq \oldweight$}
\State Assert that $\alpha_1, \alpha_2$ are the choices given by $\Hash(\newball)$

\If{$\oldweight, \newweight \in \left(\frac{2^{k-1}}{\log m},\frac{2^{k}}{\log m}\right]_\R$ for some $k$} 
\State Update the weight of $b_i$ to $\newweight$ directly
\Else  
\State \multiline{Mark $\oldball$ as \leftover ball (it is still considered as a ball of weight $\oldweight$)} 
\State $\InsertBall(\oldball, \newweight,B_{\alpha_1},B_{\alpha_2})$
\EndIf
\end{algorithmic}
\end{flushleft}
\end{algorithm}

\section{Dynamic Page Efficient $\SSE$}
\label{sec:peff_sse}
In this section, we introduce the $\SSE$ scheme $\pageSSE$ based on $\layeredchoice$. 
Essentially, we interpret lists $L_i$ of identifiers matching keyword $w_i$ as balls of a certain weight and use $\layeredchoice$ to manage the balls in $m$ bins.
Let $N$ be the maximal size of the database, $p$ be the page size and $\Hash$ be a hash function mapping into $\Int{m}^2$ for $m = \ceil{\maxweight/(\log\log\log \secpar \cdot \log\log\maxweight)}$ and $\maxweight = N/p$. 
Assume for now that $\abs{L_i} \leq p$, \ie each keyword has at most $p$ associated keywords.
Let $p\leq N^{1-1/\log\log \secpar}$.
(This is needed for the requirement $m\geq\secpar^{1/\log\log\secpar}$ of $\layeredchoice$, see \Cref{thm:layeredchoice}.)
For convenience, we adapt the notation of $\layeredchoice$ to such lists as follows\footnote{As \Cref{algo:layeredchoice} is kept purely combinatorial, balls technically have no content.
We still need to retreive lists $L$ given the keyword $w$ in this context. 
Thus, we say that the pair $(w,L)$ is a ball identified by $w$ and scaled weight $\abs{L}/p$. We assume that we can retrieve the list $L$ given $w$ from the bin that contains ball $(w,L)$. Clearly, this does not change the behaviour of $\layeredchoice$ and we can still apply \Cref{thm:layeredchoice} on the given variant.}:
\begin{itemize}
    \item $\layeredchoice.\Setup(\{(w_i,L_i)\}_{i=1}^W, \maxweight)$:
        We interpret the pair $(w_i,L_i)$ as a ball with identifier $w_i$ and weight $\abs{L_i}/p \in [0,1]$, where $L_i$ is a list of (at most $p$) identifiers matching keyword $w_i$. The bin choices for $(w_i,L_i)$ are given by $\alpha_1,\alpha_2 \gets \Hash(w_i)$.
        Run the setup defined in \Cref{algo:layeredchoice} given these balls.
    \item $\layeredchoice.\InsertBall((w,L),B_{\alpha_{1}},B_{\alpha_{2}})$:
        Insert ball $(w,L)$ into either bin $B_{\alpha_{1}}$ or bin $B_{\alpha_{2}}$ as in \Cref{algo:layeredchoice}.
    \item $\layeredchoice.\Update((w,L),L',B_{\alpha_{1}},B_{\alpha_{2}})$:
        Update the weight of ball $(w,L)$ to weight $\abs{L \addlist L'}/p$ as in \Cref{algo:layeredchoice} and add identifiers $L'$ to list $L$. One of the bins now contains the ball $(w,L\addlist L')$. If the new weight lies in a different subinterval, one bin contains a \leftover ball $(w,L)$ that we consider to not match $w$ anymore.
\end{itemize}

\subsection{$\pageSSE$}
Here, we describe the dynamic page efficient symmetric searchable encryption scheme $\pageSSE$ based on $\layeredchoice$. 
For a concise overview, we assume that $\ell_i \leq p$ and ignore delete operations for now. 
Also, we present a version of the scheme with an update that requires $2$ RTTs.
Later, we show how to treat arbitrary list sizes, introduce delete operations and show how to obtain updates in $1$ RTT.
A detailed description of $\pageSSE$ is given in \cref{algo:page_sse}.

\para{$\pageSSE.\KeyGen(1^\secpar)$} 
Sample encryption key ${\KENC}$ for $\Enc$ with the given security parameter $\secpar$.
Return the client's master secret key $\K = \KENC$.

\para{$\pageSSE.\Setup(\K,N,\DB)$} 
Receive as input the client's secret key $\K$, an upperbound $N$ on the number of identifiers  and the initial database $\DB = (\DB(w_i))_{i=1}^W$. 
Recall that $\DB(w_i) = (\ind_1, \dots, \ind_{\ell_i})$ is a list of $\ell_i$ document identifiers and that $\sum_{i=1}^W\ell_i \leq N$. 
interpret $(w_i,\DB(w_i))$ as a ball of weight $\ell_i/p \in [0,1]$ and call $\layeredchoice.\Setup$ with maximal weight $N/p$ and balls $(w_i,\DB(w_i))_{i=1}^W$ as input. The two random choices $(\alpha_{i,1},\alpha_{i,2}) \gets \Hash(w_i)$ in $\layeredchoice.\Setup$ are drawn by evaluating $\Hash$ on $w_i$.
The result are $m$ bins $(B_i)_{i=1}^m$ filled with the balls such that each bin has load at most $c\log\log\log(\secpar)\log \log (N/p)$ (see \Cref{thm:layeredchoice}).
Thus, each bin contains at most $p\cdot c\log\log\log(\secpar)\log \log (N/p)$ identifiers as weights are scaled by a factor $p$. (The constant $c\in\N$ only depends on $N$ but not the output of $\layeredchoice.\Setup$.)
Next, each bin is filled up to maximal size with dummy items.
Finally, encrypt the bins $B_i\encr \gets \Enc_{\KENC}(B_i)$ and return $\EDB = (B_i\encr)_{i=1}^m$.

\para{$\pageSSE.\Search(\K,w;\EDB)$} 
The client receives its secret key $\K$ and keyword $w$. 
She sends $w$ to the server and in return receives bins $B_{\alpha_1}\encr,B_{\alpha_2}\encr$, where $(\alpha_1,\alpha_2) \gets \Hash(w)$.

\para{$\pageSSE.\Update(\K,(w,L'), \add;\EDB)$} 
The client receives its secret key $\K$, keyword $w$ and a list $L'$ of new identifiers matching $w$. 
She sends $w$ to the server and again receives bins $B_{\alpha_1}\encr,B_{\alpha_2}\encr$ in return, where $(\alpha_1,\alpha_2) \gets \Hash(w)$.
Next, the client decrypts $B_{\alpha_1}\encr,B_{\alpha_2}\encr$ to $B_{\alpha_1},B_{\alpha_2}$ and retrieves ball $(w,L)$ from the corresponding bin $B_{\alpha}\in\{B_{\alpha_1}, B_{\alpha_2}\}$.
Then, she calls $\layeredchoice.\UpdateBall$ with old ball $(w,L)$, new identifiers $L'$ and bins $B_{\alpha_1},B_{\alpha_2}$ to insert the new identifiers $L'$ into $B_{\alpha}$.
Finally, she reencrypts the bins and sends them to the server. The server then replaces the old bins with the updated bins.

\begin{algorithm}[h]
\begin{flushleft}
\caption{$\pageSSE$}
\label{algo:page_sse}
\underline{$\pageSSE.\KeyGen(1^\secpar)$}
\begin{algorithmic}[1]
\State Sample ${\KENC}$ for $\Enc$ with security parameter $\secpar$
\State \Return $K = \KENC$
\end{algorithmic}
\underline{$\pageSSE.\Setup(\K,N,\DB)$}
\begin{algorithmic}[1]
\State \multiline{Set $B_1, \dots, B_m \gets \layeredchoice.\Setup(\{(w_i,\DB(w_i))\}_{i=1}^W, N/p)$}
\State Fill bins $B_1, \dots, B_m$ up to size $p\cdot c\log\log\log(\secpar) \log\log (N/p)$ with zeroes
\State Set $B_i\encr \gets \Enc_{\KENC}(B_i)$ for $i\in[1,m]$
\State \Return $\EDB = (B_1\encr, \dots, B_m\encr)$
\end{algorithmic}
\underline{$\pageSSE.\Search(\K,w;\EDB)$}\\
\textit{Client:}
\begin{algorithmic}[1]
\State \Return $w$
\end{algorithmic}
\textit{Server:}
\begin{algorithmic}[1]
\State Set $\alpha_{1},\alpha_{2}\gets\Hash(w)$
\State \Return $B_{\alpha_{1}}\encr, B_{\alpha_{2}}\encr$
\end{algorithmic}
\underline{$\pageSSE.\Update(\K,(w,L'), \add;\EDB)$}\\
\textit{Client:}
\begin{algorithmic}[1]
\State \Return $w$
\end{algorithmic}
\textit{Server:}
\begin{algorithmic}[1]
\State Set $\alpha_{1},\alpha_{2}\gets\Hash(w)$
\State \Return $B_{\alpha_{1}}\encr, B_{\alpha_{2}}\encr$
\end{algorithmic}
\textit{Client:}
\begin{algorithmic}[1]
\State Set $B_{\alpha_i}\gets \Dec_{\KENC}(B_{\alpha_{i}}\encr)$ for $i\in\{1,2\}$
\State Retrieve ball $(w,L)$ from $B_{\alpha}$ for appropriate $\alpha\in\{\alpha_1,\alpha_2\}$
\State Run $\layeredchoice.\UpdateBall((w,L),L',B_{\alpha_1},B_{\alpha_2})$
\State Set $B_{\alpha_i}\new\gets \Enc_{\KENC}(B_{\alpha_{i}})$ for $i\in\{1,2\}$
\State \Return $B_{\alpha_2}\new, B_{\alpha_2}\new$
\end{algorithmic}
\textit{Server:}
\begin{algorithmic}[1]
\State Replace $B_{\alpha_i}\encr$ with $B_{\alpha_i}\new$ for $i\in\{1,2\}$
\end{algorithmic}
\end{flushleft}
\end{algorithm}

\subsection{Security}
\label{sub:sec_page_sse}
The scheme $\pageSSE$ is correct as each keyword has two bins that contain its identifiers associated to it (and these bins are consistently retrieved and updated with $\layeredchoice$).
If the hash function is modeled as a random oracle, the bin choices are uniformly random and \Cref{thm:layeredchoice} guarantees that bins do not overflow.

Also, $\pageSSE$ is selectively secure and has standard setup leakage $N$, such as search and update leakage $\qpat$, where $\qpat$ is the query pattern\footnote{This is equivalent to page length hiding leakage $\LH$, as we only restrict ourselves to lists of size at most $p$.}.
This can be shown with a simple hybrid argument. We sketch the proof here and refer to \Cref{sec:peff_sec} for more details.
For setup, the simulator $\Sim$ receives $N$, recomputes $m$ and initializes $m$ empty bins $B_1,\dots,B_m$ of size $p\cdot c\log\log\log(\secpar)\log\log (N/p)$ each. $\Sim$ then outputs $\EDB' = (\Enc_{\KENC'}(B_i)_{i=1}^m)$ for some sampled key $\KENC'$. As $\Enc$ is $\INDCPA$ secure (and bins do not overflow in the real experiment except with negligible probability), the output $\EDB'$ is indistinguishable from the output of $\Setup$ in the real experiment.
For a search query on keyword $w$, $\Sim$ checks the query pattern $\qpat$ whether $w$ was already queried. If $w$ was not queried before, $\Sim$ a new uniformly random keyword $w'$. Otherwise, $\Sim$ responds with the same keyword $w'$ from the previous query. As we assume that keywords are preprocessed by the client via a $\PRF$, the keywords $w$ and $w'$ are indistinguishable.
For an update query on keyword $w$, the client output in the first flow is the same as in a search query and thus, $\Sim$ can proceed as in search. For the second flow, $\Sim$ receives two bins $B_{\alpha_1},B_{\alpha_2}$ from the adversary, directly reencrypts them and sends them back to the adversary. This behaviour is indistinguishable, as the bins are encrypted and again, bins do not overflow except with negligible probability.

For adaptive security, the adversary can issue search and update queries that depend on previous queries. As \Cref{thm:layeredchoice} assumes selectively chosen $\InsertBall$ and $\UpdateBall$ operations, there is no guarantee that bins do not overflow anymore in the real game. Thus, the adversary can potentially distinguish update queries of the simulated game from real update queries if she manages to overflow a bin in the real game, as she would receive bins with increased size only in the real game. 
Fortunately, we can just add a check in $\Update$ whether one of the bins overflows after the $\layeredchoice.\UpdateBall$ operation. In that case, the client reverts the update and send back the (reencrypted) original bins. Now, \Cref{thm:layeredchoice} still guarantees that bins overflow only with negligible probability after $\Setup$ and we can show that the simulated game is indistinguishable from the real game as before. (Note that $\pageSSE$ is still correct after this modification, since queries are chosen selectively for correctness.)
Note that when the client remarks that a bin overflowed in an $\Update$ in a real world environment, this is due malicious $\Update$ operations. The client can adapt his reaction accordingly, whereas the server learns no information about the attack without being notified by the client.

We can show that $\pageSSE$ with the adjustement of $\Update$ is correct $\LH$-adaptively secure. The same simulator $\Sim$ suffices and we omit the details.

\subsection{Extensions}

\subsubsection{Handling Long Lists.}
We now adapt $\pageSSE$ to handle arbitrary lists $L$ (with potentially more than $p$ identifiers). 
(We proceed similarly to the static scheme \textsf{Pluto} from \cite{C:BBFMR21} and extend the ideas to updates.)
For this, we split $L$ into sublists of size at most $p$. 
The (encrypted) full sublists of size $p$ can be stored in a hash table $\TFull$ on the server and the incomplete sublists are handled by $\pageSSE$ as before.
For search, the client needs to know the number of sublists in order to fetch the right amount from the server. 
This information is also required for update queries in order to know when to insert another full list into $\TFull$.
This information can be outsourced in a table $\TLen$. 
Here, the client stores for each keyword $w$ (with $\ell$ matching identifiers) the number of sublists $\TLen[w] = \ceil{\ell/p}$ in encrypted format. 
In the following, we describe the updated $\Setup,\Search$ and $\Update$ of $\pageSSE$ in more detail.

\para{Setup}
For setup, let $L_i$ be a list of $\ell_i$ identifiers matching keyword $w_i$ and $\PRF$ be a secure pseudo-random function mapping to $\{0,1\}^{\ceil{\log(N)}}$.
We set $x_i = \ceil{\ell_i/p}$. 
The client splits $L_i$ into sublists $L_{i,1},\dots,L_{i,x_i-1}$ of size $p$ and sublist $L_{i,x_i}$ of size at most $p$. 
She evaluates $m_i \gets \PRF_{\KPRF}(w_i)$, where $\KPRF$ is a key for $\PRF$ sampled in $\KeyGen$. 
The mask $m_i$ is used to encrypt the content of $\TLen$.
After initializing the table $\TLen$ with $N$ random entries of size $\log(N)$ bits and $\TFull$ with $N/p$ (arbitrary) lists of size $p$, she sets $\TLen[w_i] = x_i \xor m_i$ and $\TFull[w \conc i] = L_{i,j}$ for $j\in[1,x-1]$. 
Next, she generates $(B_i)_{i=1}^m$ as before with the incomplete lists $L_{i,x_i}$ except that the bin choices for list $L_{i,x_i}$ are $(\alpha_{i,1},\alpha_{i,2}) \gets \Hash(w_i \conc x_i)$. 
(This is because after some updates, the incomplete sublist of $w_i$ might become full and a new incomplete sublist has to be started. 
When the new incomplete sublist gets inserted with $\layeredchoice$, it is interpreted as a new ball and new bins need to be chosen.)
Finally, she encrypts the content of $\TFull$ and returns $\EDB = (\TLen, \TFull, (B_i\encr)_{i=1}^m)$.

\para{Search}
For search queries on keyword $w$, the client outputs mask $m\gets\PRF_{\KPRF}(w)$ in addition to $w$.
The server uses this mask to decrypt the number of sublists $x \gets \TLen[w] \xor m$, retreives $x - 1$ encrypted sublists $L_i \gets \TFull[w \conc i]$ from the table for $i\in[1,x-1]$ and the two bins $B_{\alpha_1}\encr$ and $B_{\alpha_2}\encr$ via $(\alpha_{1},\alpha_{2}) \gets \Hash(w \conc x)$. 
Finally, the server sends the encrypted bins and sublists to the client. 
Clearly, the client obtains all matching identifiers after decrypting the received lists and bins. 

\para{Update}
For update queries on keyword $w$ and list $L'$ of (at most $p$) new identifiers\footnote{For updates with more than $p$ identifiers, the client can use the update multiple times.}, the client generates mask $m$ as before and sends $(w,m)$ to the server. 
The server again decrypts $x$ from $\TLen$ and sends $B_{\alpha_1}\encr,B_{\alpha_2}\encr$ to the client. 
In addition, the server already sends the bins $B_{\alpha_3}\encr,B_{\alpha_4}\encr$ for $\alpha_3,\alpha_4 \gets \Hash(w \conc x+1)$ to the client (in case the incomplete list overflows).
The client now retrieves the old (incomplete) list $L$ of identifiers matching $w$ from the decrypted bins $B_{\alpha_1},B_{\alpha_2}$. We distinguish two cases:
\begin{enumerate}
    \item  If $L\addlist L'$ contains more than $p$ identifiers, the client sets $L\new = L\addlist L'$ and marks $(w,L)$ as a \leftover ball inside $B_{\alpha_1},B_{\alpha_2}$.
    Then, she splits $L\new$ into two sublists $L^{=p}$ with $p$ identifiers and $L^{\leq p}$ of at most $p$ identifiers.
    The client then inserts list $L^{\leq p}$ into bins $B_{\alpha_3},B_{\alpha_4}$ via $\layeredchoice.\InsertBall((w,L^{\leq p}), B_{\alpha_3},B_{\alpha_4})$ and sends the updated (reencrypted) bins $\{B_i\encr\}_{i=1}^4$ such as encrypted list $L\encr = \Enc_{\KENC}(L^{=p})$ to the server.
    \item Otherwise, the client proceeds as before, \ie adds the new identifieres $L'$ to ball $(w,L)$ via $\UpdateBall$ and reencrypts the received bins.
\end{enumerate}
Finally, the server replaces the old bins with the reencrypted bins, and if she received an encrypted  list $L\new$, she stores the received list in $\TFull[w \conc x+1] = L\encr$ and updates $\TLen[w] = x+1$. 

\para{Leakage profile} Now, search and update queries $\pageSSE$ clearly leak the number of sublists $x = \ceil{\ell/p}$ for a given keyword $w$ with $\ell$ matching identifiers. Further, update leaks when a list was completed. Thus, update leaks $\ceil{\ell + \abs{L'}}$. This is exaclty the leakage modeled by $\LH$.
As tables $\TLen$ and $\TFull$ are encrypted, it is straigthforward to adapt the security analysis in \Cref{sub:sec_page_sse} to the extended scheme with respect to leakage function $\LH$.

\subsubsection{Handling Deletes.}
We apply the generic solution from \cite{CCS:Bost16} to handle deletes. We use two instantiations of $\pageSSE$, $\Sigma_\add$ for added items and $\Sigma_\del$ one for deletes. 
For adding identifiers $L'$ to a keyword $w$, the client adds list $L'$ to $\Sigma_\add$ via $\Sigma_\add.\Update(\K,(w,L'),\add;\EDB)$.
For deleting identifiers $L'$ from a keyword $w$, the client adds list $L'$ to $\Sigma_\del$ via $\Sigma_\del.\Update(\K,(w,L'),\add;\EDB)$.
For a search query, the client fetches the identifiers $w$ from both $\Sigma_\add$ and $\Sigma_\del$ and removes the set of items $L_\del$ received from $\Sigma_\del$ from the set of items $L_\add$ received from $\Sigma_\add$, \ie sets $L \gets L_\add \setminus L_\del$.

\subsubsection{Optimized RTT.}
Search queries of $\pageSSE$ need only $1$ RTT, whereas update queries unfortunately require $2$ RTTs. 
We can use \enquote{piggybacking} in order to reduce the update RTT to $1$ as follows.
Instead of sending the second flow of the update query directly to the server, the client stashes the response and waits for the next query (either update or search). 
On the next query, the client sends the stashed response in addition to the query. 
The server then finishes the pending update query (by storing the received bins and updating the tables) and responds the query subsequently.

\subsection{Efficiency}
We now inspect the efficiency of $\pageSSE$.
Let $\maxweight = N/p$.
The server stores $m = \ceil{\maxweight/(\log\log\log \secpar \cdot \log\log\maxweight)}$ bins of size $\bigO{p\log\log\log(\secpar)\log\log(\maxweight)}\cdot \bigO{\secpar}$ each, tables $\TLen$ with $N$ entries of size $\log(N)$ and $\TFull$ with $N/p$ entries of size $p\cdot\bigO{\secpar}$ each.
(Recall that a single identifier has size $\bigO{\secpar}$.)
As $N = \poly$, the storage efficiency is $\bigO{1}$ in total. 
There is no client stash required\footnote{The version of $\pageSSE$ with $1$ RTT updates requires a stash of size $\bigOtilde{p\log\log (N/p)}$ to temporarily store the second flow of the update query until the next query.}.
Further, the server looks up $4$ bins of capacity $\bigOtilde{p\log\log(N/p)}$ and $x-1$ encrypted lists of $p$ identifiers from $\TFull$ for a search query on word $w$, where $x$ is the number of pages needed to store the document indices matching keyword $w$ in plaintext. 
Thus, the page efficiency is $\bigOtilde{\log\log\frac{N}{p}}$.
This further implies that $\pageSSE$ has $\bigO{1}$ locality if only lists up to size $p$ are inserted.

\begin{theorem}[$\pageSSE$]
\label{thm:peff_sec}
Let $N$ be an upper bound on the size of database $\DB$ and $p$ be the page size. 
Let $p \leq N^{1-1/\log\log \secpar}$.
The scheme $\pageSSE$ is correct and $\LH$-adaptively semantically secure if $\Enc$ is $\INDCPA$ secure and $\Hash$ is modeled as a random oracle.
It has constant storage efficiency and $\bigOtilde{\log\log N/p}$ page efficiency.
If only lists up to size $p$ are inserted, $\pageSSE$ has constant locality.
\end{theorem}
\begin{proof}
Efficiency and security follow from the discussions above.
\end{proof}
\section{The Generic Local Transform}

\subsection{Preliminaries}
\label{sec:OSSE}

{\bf Suitable page-efficient SSE.}
The interface of the page-efficient scheme used within the Generic Local Transform extends the standard SSE interface defined in \Cref{sec:prelim}, in two ways.
\begin{itemize}
\item First, $\Setup(N,p,\DB)$ takes a new argument as input: the page size $p$. The transform will create many instances of the underlying page-efficient scheme, each with a different page size. This makes it necessary to specify the page size during setup.
\item Second, in the $\Update(w,S)$ procedure, the second parameter $S$ is a \emph{set} of document identifiers. The correctness requirement is that all identifiers in $S$ should be added to the list for keyword $w$.
The standard definition of $\Update$, where a single identifier is added, corresponds to the case where $S$ is a singleton. $S$ is allowed to be empty, in which case nothing is added.
\end{itemize}

If a scheme instantiates that interface, and, in addition, satisfies the following three conditions, we will call such as scheme a \emph{suitable} page-efficient SSE.
\begin{itemize}
\item The scheme has client storage $\bigO{1}$.
\item The scheme has locality $\bigO{1}$ during searches and updates \emph{when accessing a list of length at most one page}.
\item The leakage of the scheme is page-length-hiding.
\end{itemize}

\medskip
\noindent
{\bf Overflowing SSE.}
We introduce the notion of \emph{Overflowing} SSE. An Overflowing SSE (OSSE) has the same interface and functionality as a standard SSE scheme, except that during a $\Setup$ or $\Update$ operation, it may refuse to store some document identifiers. Those identifiers are called \emph{overflowing}. At the output of the $\Setup$ and $\Update$ operations, the client returns the set of overflowing elements. Compared to standard SSE, the correctness definition is relaxed in the following way: during a $\Search$, only matching identifiers that were \emph{not} overflowing need to be retrieved.

The intention of an Overflowing SSE is that it may be used as a component within a larger SSE scheme, which will store the overflowing identifiers using a separate mechanism. The use of an OSSE may be regarded as implicit in some prior SSE constructions. We have chosen to introduce the notion explicitly because it allows to cleanly split the presentation of the Generic Local Transform into two parts: an OSSE scheme that stores most of the database, and an array of page-efficient schemes that store the overflowing identifiers.

\subsection{Dynamic Two-Dimensional One-Choice Allocation}

The first component of the Generic Local Transform is an OSSE scheme, $\clippedOSSE$.
In line with prior work, we split the presentation of $\clippedOSSE$ into two parts: an allocation scheme, which specifies where elements should be stored; and the SSE scheme built on top of it, which adds a layer of encryption, key management, and other mechanisms needed to convert the allocation scheme into a full SSE.

The allocation scheme within $\clippedOSSE$ is called $\onechoicealloc$. Similar to \cite{STOC:ANSS16}, the allocation scheme is an abstract construct that defines the memory locations where items should be stored, but does not store anything itself. In the case of $\onechoicealloc$, items are stored within buckets, and the procedures return as output the indices of \emph{buckets} where items should be stored. From the point of view of $\onechoicealloc$, each bucket has unlimited storage. In more detail, $\onechoicealloc$ contains two procedures, $\Fetch$ and $\Append$.
\begin{itemize}
\item[--] $\Fetch(m,w,\ell)$: given a number of buckets $m$, a keyword $w$, and a list length $\ell$, $\Fetch$ returns (a superset of) the indices of buckets where elements matching keyword $w$ may be stored, assuming there are $\ell$ such elements.
\item[--] $\Append(m,w,\ell)$: given the same input, $\Append$ returns the index of the bucket where the \emph{next} element matching keyword $w$ should be inserted, assuming there are currently $\ell$ matching elements.
\end{itemize}
The intention is that $\Append$ is used during an SSE $\Update$ operation, in order to choose the bucket where the next list element is stored; while $\Fetch$ is used during a $\Search$ operation, in order to determine the buckets that need to be read to retrieve all list elements.  $\onechoicealloc$ will satisfy the correctness property given in \Cref{def:onechoicecorrect}.
Note that the number of buckets $m$ is always assumed to be a power of 2.

\begin{definition}[Correctness]
\label{def:onechoicecorrect}
For all $m$, $w$, $\ell$, if $m$ is a power of 2, then
\[
\bigcup_{0\leq i \leq \ell-1} \Append(m,w,i) \subseteq \Fetch(m,w,\ell).
\]
\end{definition}

To describe $\onechoicealloc$, it is convenient to conceptually group buckets into superbuckets. For $\ell = 2^i \leq m$, an \emph{$\ell$-superbucket} is a collection of $\ell$ consecutive buckets, with indices of the form $k\cdot \ell, k\cdot \ell +1, \dots, (k+1)\cdot \ell -1$, for some $k \leq m/\ell$. A $1$-superbucket is the same as a bucket. Notice that for a given $\ell$, $\ell$-superbuckets do not overlap. They form a partition of the set of buckets. For $\ell > 1$, each $\ell$-superbucket contains exactly two $\ell/2$-superbuckets.

Let $H$ be a hash function, whose output is assumed to be uniformly random in $\{1,\dots,m\}$.
$\onechoicealloc$ works as follows.
Fix a keyword $w$ and length $\ell \leq m$ (the case $\ell > m$ will be discussed later).
Let $\ell' = 2^{\lceil \log \ell \rceil}$ be the smallest power of 2 larger than $\ell$.
On input $w$ and $\ell$, $\onechoicealloc.\Fetch$ returns the (unique) $\ell'$-superbucket that contains $H(w)$.

\begin{algorithm}[h]
\caption{Dynamic Two-Dimensional One-Choice Allocation (\onechoicealloc)}
\label{algo:onechoicealloc}
\begin{minipage}[t]{0.5\textwidth}
\underline{$\onechoicealloc.\Fetch(m,w,\ell)$}
\begin{algorithmic}[1]
\State $\ell' \gets 2^{\lceil \log \ell \rceil}$
\If{$\ell' \geq m$}
\State \Return $\{0, \dots, m-1\}$
\Else
\State $i \gets \lfloor H(w)/{\ell'} \rfloor$
\State  \Return $\{{\ell'}\cdot i, \dots, {\ell'}\cdot i + {\ell'}-1\}$
\EndIf
\end{algorithmic}
\end{minipage}%
\begin{minipage}[t]{0.5\textwidth}
\underline{$\onechoicealloc.\Append(m,w,\ell)$}
\begin{algorithmic}[1]
\State $\ell \gets \ell \bmod m$
\State $\ell' \gets 2^{\lceil \log (\ell + 1) \rceil}$
\State $i \gets \lfloor H(w)/{\ell'} \rfloor$
\If{$\lfloor 2H(w)/{\ell'} \rfloor \bmod 2 = 0$}
\State  \Return ${\ell'}\cdot i + \ell$
\Else
\State  \Return ${\ell'}\cdot i + \ell - {\ell'}/2$
\EndIf
\end{algorithmic}
\end{minipage}
\end{algorithm}

Meanwhile, $\onechoicealloc.\Append$ is designed in order to ensure that the first $\ell$ successive locations returned by $\Append$ for keyword $w$ are in fact included within the $\ell'$-superbucket above $H(w)$ (that is, in order to ensure correctness). For the first list element (when $\ell = 0$), $\Append$ returns the bucket $H(w)$; for the second element, it returns the other bucket contained inside the 2-superbucket above $H(w)$. More generally, if $S$ is the smallest superbucket above $H(w)$ that contains at least $\ell+1$ buckets, $\Append$ returns the leftmost bucket within $S$ that has not yet received an element. In practice, the index of that bucket can be computed easily based on $\ell$ and the binary decomposition of $H(w)$, as done in \Cref{algo:onechoicealloc}.
(In fact, the exact order in which buckets are selected by $\Append$ is irrelevant, as long as it selects distinct buckets, and correctness holds.)

When the size of the list $\ell$ grows above the number of buckets $m$, $\Fetch$ returns all buckets, while $\Append$ selects the same buckets as it did for $\ell \bmod m$.


\subsection{Clipped One-Choice OSSE}

$\clippedOSSE$ is the OSSE scheme obtained by storing lists according to $\onechoicealloc$, using $m = O(N/\log \log N)$ buckets, with each bucket containing up to $\tau = \lceil \alpha \log \log N\rceil$ items, for some constant $\alpha$.
Buckets are always padded to the the threshold $\tau$ and encrypted before being stored on the server. Thus, from the server's point of view, they are completely opaque.
A table $T$ containing (in encrypted form) the length of the list matching each keyword $w$ is also stored on the server.

Given $\onechoicealloc$, the details of $\clippedOSSE$ are straightforward. A short overview is given in text below.
The encrypted database generated by $\Setup$ is essentially equivalent to starting from an empty database, and populating it by making repeated calls to $\Update$, one for each keyword--document pair in the database. For that reason, we focus on $\Search$ and $\Update$.
The full specification for $\Setup$, $\Search$, and $\Update$ is given as pseudo-code in \Cref{algo:clippedOSSE}.

\begin{itemize}
\item{$\Search$:} to retrieve the list of identifiers matching keyword $w$, $\clippedOSSE$ calls $\onechoicealloc(m,w,\ell)$ to get the set of bucket indices where the elements matching keyword $w$ have been stored.
The client retrieves those buckets from the server, and decrypts them to obtain the desired information.
\item{$\Update$:} to add a new item to the list matching keyword $w$, $\clippedOSSE$ calls $\onechoicealloc(m,w,\ell)$ to determine the bucket where the new list item should be inserted.
The client retrieves that bucket from the server, decrypts it, adds the new item, reencrypts the bucket, and sends it back to the server.
If that bucket was already full, the item is overflowing, in the sense of \Cref{sec:OSSE}.
\end{itemize}

\begin{algorithm}[h]
\caption{Clipped One-Choice OSSE $(\clippedOSSE)$}
\label{algo:clippedOSSE}
Global parameters: constants $d, \alpha \in \N^*$\\
\begin{minipage}[t]{0.5\textwidth}
\underline{$\clippedOSSE.\KeyGen$}
\begin{algorithmic}[1]
\State Generate key $K$, $K_\PRF$ for $\Enc$, $\PRF$
\State \Return $K, K_\PRF$
\end{algorithmic}
\underline{$\clippedOSSE.\Setup(N,\DB)$}
\begin{algorithmic}[1]
\State $m \gets  2^{\lceil \log(N/\log \log N)\rceil}$
\State $\tau \gets \lceil \alpha \log \log N \rceil$
\State $B_0, \dots, B_{m-1}, T, \EDB, \clip \gets \varnothing$
\ForAll{each $(w,\{e_1,\dots,e_\ell\})$ in $\DB$}
\State $K_w \gets \PRF_{K_\PRF}(w)$
\State $T[w] \gets \Enc_{K_w}(\ell)$
\ForAll{$t$ from $1$ to $\ell$}
\State $C \gets \varnothing$
\State $i \gets \onechoicealloc.\Append(m,w,t-1)$
\If $|B[i]| < \tau$
\State $B[i] \gets B[i] \cup \{e_i\}$
\Else
\State $C \gets C \cup \{e_i\}$
\EndIf
\If{$|S| > 0$}
\State $\clip \gets \clip \cup (w,\ell,C)$
\EndIf
\EndFor
\EndFor
\State Let $B^\enc[i] = \Enc_K(B_i)$ for each $i$
\State \Return $\EDB = (T, (B^\enc[i])), \clip$
\end{algorithmic}
\end{minipage}%
\begin{minipage}[t]{0.5\textwidth}
\underline{$\clippedOSSE.\Search(w)$}
\begin{algorithmic}[1]
\Statex \emph{Client (Search token):}
\State \Send $w, K_w = \PRF_{K_\PRF}(w)$
\Statex \emph{Server:}
\State $\ell \gets \Dec_{K_w}(T[w])$
\State $S \gets \onechoicealloc.\Fetch(m,w,\ell)$
\State \Return $\{B^\enc[i] : i \in S\}$
\end{algorithmic}
\underline{$\clippedOSSE.\Update(w,e)$}
\begin{algorithmic}[1]
\Statex \emph{Client (Update token):}
\State \Send $w, K_w = \PRF_{K_\PRF}(w)$
\Statex \emph{Server:}
\State $\ell \gets \Dec_{K_w}(T[w])$
\State $i \gets \onechoicealloc.\Append(m,w,\ell)$
\State \Send $B^\enc[i]$
\Statex \emph{Client:}
\State $B \gets \Dec_K(B^\enc[i])$
\If{$|B| < \tau$}
\State $\clip \gets \varnothing$
\State $B \gets B \cup \{e\}$
\Else
\State $\clip \gets \{e\}$
\EndIf
\State \Send $B' = \Enc_K(B)$
\Statex \emph{Server:}
\State $B^\enc[i] \gets B'$
\Statex \emph{Client:}
\State \Return $\clip$
\end{algorithmic}
\end{minipage}
\end{algorithm}

\subsection{The Generic Local Transform}
\label{subsec:GLT}

The Generic Local Transform takes as input a length-hiding page-efficient SSE scheme $\peff$.
It outputs a local SSE scheme $\local{\peff}$.

To realize $\local{\peff}$, we use two structures. The first structure is an instance of $\clippedOSSE$, which stores most of the database.
The second structure is an array of $\nlevel$ instances of $\peff$. The $i$-th instance, denoted $\peff_i$, has page size $2^i$. The $\peff_i$ instances
are used to store elements that overflow from $\clippedOSSE$.
In addition, a table $T$ stores (in encrypted form) the length of the list matching keyword $w$, for each keyword\footnote{The same table exists in $\clippedOSSE$. In an actual implementation, they would be the same table, but using $\clippedOSSE$ in black box eases the presentation.}.

Fix a keyword $w$, matching $\ell$ elements.
Let $\ell' = 2^{\lceil \log \ell \rceil}$ be the smallest power of 2 larger than $\ell$. Let $i = \log \ell'$.
At any point in time, the elements matching $w$ are stored in two locations: $\clippedOSSE$, and $\peff_i$.
Each of these two locations stores part of the elements: $\clippedOSSE$ stores the elements that did not overflow, and $\peff_i$ stores the overflowing elements.
Each element exists in only one of the two locations.

\begin{algorithm}[h]
\caption{Generic Local Transform ($\local{\peff}$)}
\label{algo:generic}
Global parameters: constant $d \in \N^*$\\
\begin{minipage}[t]{0.5\textwidth}
\underline{$\local{\peff}.\KeyGen$}
\begin{algorithmic}[1]
\State Generate key $K_\PRF$ for $\PRF$
\State \Return $K, K_\PRF$
\end{algorithmic}
\underline{$\local{\peff}.\Update(w,e)$}
\begin{algorithmic}[1]
\Statex \emph{Client (Update token):}
\State \Send $w, e, K_w = \PRF_{K_\PRF}(w)$
\Statex \emph{Server:}
\State $C \gets \clippedOSSE.\Update(w,e)$
\State $\ell \gets \Dec_{K_w}(T[w])$
\State $T[w] \gets \Enc_{K_w}(\ell+1)$
\State \Send $\ell$
\Statex \emph{Client:}
\If{$\lceil \log \ell \rceil = \lceil \log (\ell+1) \rceil$}
\State $\peff_{\lceil \log \ell\rceil}.\Update(w,C)$
\Else
\State $i \gets \lceil \log \ell\rceil$
\State $S \gets$ set of matches in
\Statex \hspace{1.33cm} $\peff_i.\Search(w)$
\State $\peff_{i+1}.\Update(w,S \cup C)$
\EndIf
\end{algorithmic}
\end{minipage}%
\begin{minipage}[t]{0.5\textwidth}
\underline{$\local{\peff}.\Setup(N,\DB)$}
\begin{algorithmic}[1]
\State $\nlevel \gets \lceil N/\log^d N \rceil$
\ForAll{$(w,S) \in \DB$}
\State $K_w \gets \PRF_{K_\PRF}(w)$
\State $T[w] \gets \Enc_{K_w}(|S|)$
\EndFor
\State $\EDB, \clip \gets \clippedOSSE.\Setup(\DB)$
\ForAll{$i$ from $0$ to $\nlevel$}
\State $\DB_i \gets \{(w,C) : (w,\ell,C) \in \clip$
\Statex \hspace{1.77cm} and $2^{i-1} < \ell \leq 2^i\}$
\State $\peff_i \gets \peff.\Setup($
\Statex \hspace{2.3cm}$\lceil N/\log N \rceil, 2^i, \DB_i)$
\EndFor
\end{algorithmic}
\underline{$\local{\peff}.\Search(w)$}
\begin{algorithmic}[1]
\Statex \emph{Client (Search token):}
\State \Send $w, K_w = \PRF_{K_\PRF}(w)$
\Statex \emph{Server:}
\State $i \gets \lceil \log(\Dec_{K_w}(T[w])) \rceil$
\State \Return $\clippedOSSE.\Search(w)$
\Statex \hspace{.8cm} $\cup\; \peff_{i}.\Search(w)$
\end{algorithmic}
\end{minipage}
\end{algorithm}

\begin{itemize}
\item{$\Search$.}
During a $\Search$ operation, $\local{\peff}$ queries both structures, and combines their output to retrieve all matching elements.
\item{$\Update$.}
During an $\Update$ operation to add element $e$, $\local{\peff}$ forwards the update query to $\clippedOSSE$, and gets as output $C = \varnothing$ if the element did not overflow, or $C = \{e\}$ if the element did overflow.
For now, assume that $\lceil \log \ell \rceil = \lceil \log (\ell+1) \rceil $, that is, the $\peff_i$ instance associated with the list remains the same during the update operation.
In that case, $\peff_i$ is updated for the set $C$. (Recall from \Cref{sec:OSSE} that a length-hiding SSE such as $\peff$ accepts sets of elements as input in $\Update$.)
The length-hiding property is designed to guarantee that the content of $C$ (including whether it is empty) is not leaked to the server.
Now assume $\lceil \log \ell \rceil < \lceil \log (\ell+1) \rceil $. In that case, the $\peff$ instance associated with the list becomes $\peff_{i+1}$ instead of $\peff_i$. The client retrieves all current overflowing elements from $\peff_i$, adds the content of $C$, and stores the result in $\peff_{i+1}$.
\end{itemize}

\subsection{Overflow of $\clippedOSSE$}

The main technical result in this section is regards the number of overflowing items in $\clippedOSSE$.

\begin{theorem}
\label{thm:clippedOSSE}
Suppose that $\clippedOSSE$ receives as input a database of size $N$, such that the size of the longest list is $\bigO{N/\polylog N}$.Then for any constant $d$, there exists a choice of parameters of $\clippedOSSE$ such that the number of overflowing items is $\bigO{N/\log^d N}$.
\end{theorem}

The proof of \Cref{thm:clippedOSSE} is given in \Cref{sec:clippedOSSE}. An overview of the proof is not included here for space reasons, but can be found in \Cref{sec:clippedOSSE}, alongside the full proof.

The \GLT itself uses standard SSE techniques, and its properties follow from previous discussions.
We provide a formal statement below.

\begin{theorem}[\GLT]
\label{thm:GLT}
Let $N$ be an upper bound on the size of database $\DB$.
Suppose that $\peff$ is a  \emph{suitable} page-efficient scheme with page efficiency $P$ and storage efficiency $S$.
Then $\local{\peff}$ is a correct and secure SSE scheme with storage efficiency $\bigO{S}$, locality $\bigO{1}$, and read efficiency $P + \bigOtilde{\log\log N}$.
\end{theorem}

\nottoggle{anon}{

\section*{Acknowledgments}

The authors would like to thank Raphael Bost for his helpful comments. This work was supported by the ANR JCJC project SaFED.

}{}

\bibliographystyle{bibstyle/bibstyle}
\bibliography{cryptobib/abbrev0,biblio_aux,cryptobib/crypto}

\clearpage
\appendix


\section{Unconditional Static Local SSE}
\label{sec:appl}

\label{sec:useless}

In this section, we present our unconditional SSE scheme $\uselessSSE$ with constant locality, constant storage efficiency and $\bigO{\log^\eps N}$ read efficiency for any $\eps > 0$. 
For this, we first present a local ORAM construction in \Cref{sub:local_oram}. 
Then we construct a static SSE scheme with $\bigO{\log^\eps \secpar}$ page efficiency in \Cref{sub:stashless} that works for large page sizes and has constant client storage.
Finally, we use those schemes in order to construct $\uselessSSE$ in \Cref{sub:uncond_scheme}

\subsection{Local ORAM}
\label{sub:local_oram}
Let $c\in\N$ be arbitrary.
We now construct an ORAM $\localORAM$ with amortized constant locality and $\bigO{\orambs \cdot n^{1/c}\log^2(n)}$ bandwith, where $n$ is the size of the memory array and $\orambs = \Omega(n^{(c-1)/c})$ is the block size.
As $c$ is arbitrary, we can instantiate $\localORAM$ with bandwith overhead $\bigO{n^\eps}$ for any $\eps > 0$ and constant locality, if the block size is sufficiently large.
Our scheme follows the blueprint of the scheme of \cite{C:DemPapPap18}. 
The reader may find it helpful to refer to their scheme first.

\subsubsection{More Preliminaries.}
Before detailing the construction, we introduce some additional preliminaries.

\begin{definition}[ORAM]
\begin{itemize}
    \item $\ORAMInit(1^\secpar, \Mem)$: Client take as input security parameter $\secpar$ such as memory array $\Mem$ of $n$ values $\{(i,v_i)\}_{i=1}^n$ of $\bigO{\secpar}$ bits each. 
    Outputs client state $\st$ and encrypted memory $\EMem$.
    \item $\ORAMAccess(\st, i; \EMem)$: The client takes as input its state and index $i$. 
    The server takes as input encrypted memory $\EMem$. 
    Outputs value $v_i$ assigned to $i$ and updated state $\st$ to the client such as updated encrypted memory $\EMem'$ to the server.
\end{itemize}
\end{definition}

We require read-only ORAM with zero-failure probability for our construction. 
We say that an ORAM scheme is correct, if for any access sequence on block $i$, the retrieved block via $\ORAMAccess$ is $(i,v_i)$.
We say that an ORAM scheme is (adaptively) secure, if for any two (adaptively chosen) access sequences $S_1$ and $S_2$ of the same length, their access patterns $A(S_1)$ and $A(S_2)$ are computationally indistinguishable by anyone but the client.
We refer to \cite{C:DemPapPap18} for formal definitions.

\begin{lemma}[Local Oblivious Sort \cite{goodrich2011data,goodrich2011privacy,C:DemPapPap18}]
\label{lemma:obl_sort}
Given an array $X$ containing $n$ comparable elements, we can sort $X$ with a data-oblivious external-memory protocol $\OblSort$ that uses $\bigO{\frac{n}{b}\log^2\frac{n}{b}}$ I/O operations and local memory of $4b$ chunks, where an I/O operation is defined as the read/write of $b$ consecutive chunks of $X$.
\end{lemma}
We set chunk size $b=n^{1/c}\log^2 n$ in our ORAM scheme. This suffices for $O(1)$ locality. 
We implicitly assume that the sorted array is reencrypted (under the same encryption key as the input array).

\subsubsection{The Scheme.}
We now describe our construction $\localORAM$ of a read-only ORAM (based on \cite{C:DemPapPap18}). Let $c$ be a constant. We write $\orambs$ for the ORAM blocksize. Essentially, $\localORAM$ is a hierarchical ORAM with $c$ levels. For $n$ blocks of memory, it has constant locality and a bandwith of $\bigO{\orambs \cdot n^{1/c}\log^2(n)}$ with $\bigO{\orambs \cdot n^{1/c}\log^2(n)}$ temporary client storage, if $\orambs = \Omega(n^{(c-1)/c})$. Now, we give an overview of the construction. A detailed description is given in \Cref{algo:local_oram}.

\para{$\localORAM.\ORAMInit(1^\secpar, \Mem)$}
The client receives memory $\Mem = \{k,v_k\}_{k=1}^n$ with blocksize $\abs{k,v_k} = \orambs$. 
She allocates $c$ arrays $A_1,A_2,\dots,A_c$ with space for $n_i$ blocks each, where $n_1 = n^{1/c}$ and $n_i = n^{i/c} + n^{(i-1)/c}$.
Let $\pi_i : [1,n_i] \mapsto [1,n_i]$ be pseudorandom permutations. 
Initially, the client stores all blocks $(k,v_k)$ in $A_c$ at position $\pi_{c}(k)$.
Later on, blocks will also be stored in other levels. 
Note that while $A_c$ can hold all blocks, lower levels $A_i$ can only store up to $n^{i/c}$ blocks (and the remaining space is reserved for dummy queries). 
We would still like to store block $(k,v_k)$ at a pseudorandom position.
For this, we initialize tables $\T_i$, $i\in[2,c-1]$, which store a scaled index $\T_i[k] \in [1,n^{i/c}]$ for block $k$ (if the block is stored in level $i$). 
(Note that we do not require table $\T_1$ for the first level, as the client always retrieves the entire array $A_1$ each read and thus, no pseudorandom accesses is required for the first level.)
The block $k$ is later stored at location $\pi_i(\T_i[k])$.
Further, the client initializes sets $R_i$ that store the blocks mapped to level $A_i$. 
Initially, $R_c = \Mem$ and the other sets are empty. 
Levels are later rebuilt (with a new pseudorandom permutation) after a certain number of reads such that each item is only accessed once per level before the next rebuild (and thus resemble a random access to $A_i$). 
The client keeps track of the number of reads $\cnt_i$ at level $i$ after the last rebuild, initialized to $0$.
The client finally encrypts $A_i, R_i$ and $\T_i$ and sends them to the server and stores the encryption key, $\cnt_i$, and $\pi_i$ in its state. 

\para{$\localORAM.\ORAMAccess(\st, k; \EMem)$}
The client receives index $k$ of the block to be retrieved and its state $\st$ which she parses as $\{\pi_i\}_{i=2}^c, \{\cnt_i\}_{i=2}^{c}, \KENC$. 
The server receives the encrypted memory $\EMem$ which she parses as $(\{A_i\encr\}_{i=1}^{c}\allowbreak,\{R_i\encr\}_{i=2}^{c-1},\allowbreak \{\T_i\encr\}_{i=2}^{c-1})$.
First, the client increments the counts $\cnt_i$.
Next, the client retrieves tables $\T_i\encr$ and array $A_1\encr$ from the server\footnote{Downloading all encrypted $\T_i$ incurs a bandwith of $\mathcal{O}{(n^{(c-1)/c}\cdot \log(n))}$. As we require $\beta = \Omega({n^{(c-1)/c}})$ this cost vanishes in the total access bandwith.}.
After decryption, the client looks for the first level $A_{i_*}$ in which $(k,v_k)$ is stored. This level is the first for which the table $\T_i$ has a non-zero entry $\T_i[k]\neq 0$.
She performs a dummy query for all other levels $A_i$ (by accessing a random and unqueried position in $A_i$ using $\pi_i$) and retrieves $(k,v_k)$ from level $A_{i_*}$ (either by scanning $A_1$ if $i_* = 1$ or from $A_{i_*}$ at position $\pi_i(\T_i[k])$) with the help of the server.
Next, the client writes $(k,v_k)$ to $A_1$. 
Note that later, $A_1$ will be merged with upper levels and thus, we already prepare $\T_i[k] = \cnt_{i+1}$ and add an encryption of block $(k,v_k)$ to the set $R_i$ of blocks stored in the $i$-th level.
Last, the client rebuilds (some of) the levels if necessary. 
For this, she takes the highest $i^*$ such that $\cnt_{i^*} > n^{({i^*}-1)/c}$. 
If $i^* \geq 2$, all levels $A_i$ below $A_{i^*}$ are emptied and filled with dummy blocks.
Then, she chooses new pseudorandom permutations $\pi_i$ for $i \in [2,i^*]$ and merges all blocks from the lower levels into $A_{i^*}$ via a local oblivious sort $\OblSort$ (see \Cref{lemma:obl_sort}). 
Concretely, the client and server interactively sort $R_{i^*}$ with respect to the new $\pi_{i^*}$. 
The array $R_{i^*}$ is temporarily filled up to $n_{i^*}$ blocks with zeros in order to obtain array $A_{i^*}$ of size $n_{i^*}$. 
For the oblivious sort, we choose a chunk size of $n^{1/c}\log^2 n$ (which is sufficient for $\bigO{1}$ locality). 
Lastly, the server and client empty the lower levels (and its auxiliary data structures) $A_i,R_i,T_i$ for $i \in [2,i^*-1]$ and $A_1$. 
(Note that an empty and encrypted array $A_i$ can be constructed iteratively with the same method as $R_i$.)
Finally, the client updates its state, reencrypts the received data structures, and sends them back to the server (who updates $\EMem$ accordingly). 

\begin{theorem}[$\localORAM$]
Let $n$ be the size of the memory array, $\orambs = \Omega(n^{\frac{c-1}{c}})$ the blocksize.
The scheme $\localORAM$ is correct and secure, if the $\pi_i$'s are secure pseudorandom permutations and $\Enc$ is an $\INDCPA$ secure encryption scheme. Further, $\localORAM$ has amortized constant locality and bandwith of $\bigO{\orambs \cdot n^{1/c}\log^2(n)}$, and requires $\bigO{\orambs \cdot n^{1/c}\log^2(n)}$ temporary client storage (during a rebuild). The client state $\st$ has constant size. 
\end{theorem}
\begin{proof}
We show that scheme $\localORAM$ is correct (1), secure (2) and analyze its efficiency (3).

(1)
We need to show that for all indices $k\in[1,n]$, the server retrieves the block $(k,v_k)$ via the access protocol when executing an adaptive access sequence after the initialization.
For this, we observe that $(k,v_k)$ is either in $A_1$, $A_i[\pi_i(\T_i[k])]$ (for the minimal $i\in[2,c-1]$ such that $\T_i[k] \neq 0$) or $A_c[\pi_c(c)]$ (if no such $i$ exists). 
This is because if $(k,v_k)$ has been accessed in the last $n^{1/c}$ operations, the block will be stored in $A_1$. 
If $(k,v_k)$ has been accessed in the last $n^{i/c}$ operations but not in the last $n^{(i-1)/c}$ operations, it was shuffled into array $A_i$ at position $\T_i[k]$ during a previous rebuild of level $A_i$. Also, as it has not been accessed recently, we have $\T_j[k] = 0$ for $j<i$ as tables below are emptied during a rebuild.
Otherwise, it was never accessed before and is located in $A_c$ at initial position $\pi_c(k)$.
These values are retrieved by the client and thus, the scheme is correct.

(2)
We give an simulator $\Sim$. 
For initialization, the simulator receives $\abs{\Mem}$, the block size $\orambs$ and the security parameter $\secpar$. 
$\Sim$ outputs $\localORAM.\ORAMInit\allowbreak(1^\secpar, \Mem')$ for $\Mem' = \{i,0\}_{i=1}^n$ where the zeros are of size $\orambs$.
Under the $\INDCPA$ security, this output is indistinguishable from the real game, as the output is encrypted.
For simulating an access, $\Sim$ retrieves $A_1$ and the tables $T_i$ from the server. First, $\Sim$ increments $\cnt_i$. 
Then, $\Sim$ outputs $r_i\gets \Enc_{\KENC}(0)$ and random indices $\indA_i$ that were not yet queried in $A_i$ (since $A_i$ was last emptied).
Lastly, $\Sim$ and the server rebuild the largest level $A_i$ if $\cnt_i \geq n^{(i-1)/c}$ for some $i\geq 2$. For this, $\Sim$ simply checks (and updates) $\cnt_i$ accordingly and simulates the oblivious search with the server.
It follows by inspection that if $\pi_i$ are pseudorandom permutations and $\Enc$ is $\INDCPA$ secure, the interaction with $\Sim$ is indistinguishable from the game.

(3)
As $c$ is a constant, the client state is $\bigO{1}$.
Further, the client requires $\bigO{n^{1/c}\log^2 n}$ blocks of temporary storage for the oblivious sort (see \Cref{lemma:obl_sort}). 
We now inspect the bandwith and locality. Over the course of $n$ accesses, the following holds for the $i$-th access.
\begin{itemize}
    \item The client reads array $A_1$ of size $\bigO{\orambs\cdot n^{1/c}}$, one block from each other level $A_i$ and tables $\T_i$ of size $\bigO{n^{(c-1)/c}\cdot \log(n)} = \bigO{\orambs \cdot \log n}$, as $\beta = \Omega({n^{(c-1)/c}})$. 
    In total, this incurs $\bigO{\orambs\cdot n^{1/c}}$ bandwith and $\bigO{1}$ locality.
    \item If $i \mod n^{(i-1)/c} = 0 \wedge i\geq 2$, the client performs a rebuild of array $A_i$ via an oblivious sort with chunk size $n^{1/c}\log^2 n$. 
    According to \Cref{lemma:obl_sort} and as $A_i$ contains $\bigO{n^{i/c}}$ blocks, the sort requires $\bigO{n^{(i-1)/c}}$ I/O operations.
    In total, this incurs $\bigO{\orambs \cdot n^{i/c}\log^2 n}$ bandwith and $\bigO{n^{(i-1)/c}}$ locality and happens $n^{(c-i+1)/c}$ times during $n$ accesses.
\end{itemize}
In total, the amortized bandwith $B$ and locality $L$ are  
\begin{align*}
    L = &\frac{n\bigO{\orambs\cdot n^{1/c}} + \sum_{i=2}^{c} n^{(c-i+1)/c} \bigO{\orambs \cdot n^{i/c}\log^2 n}}{n} = \bigO{\beta \cdot n^{1/c} \log^2 n},\\
    B = &\frac{n\bigO{1}+\sum_{i=2}^c n^{(c-i+1)/c}\bigO{n^{(i-1)/c}}}{n} = \bigO{1}.\qedhere
\end{align*}
\end{proof}

\begin{remark}[On Deamortization]
Generally, hierarchical ORAMs can be deamortized by continuously reshuffling the layers each operation \cite{goodrich2011oblivious}.
Indeed, our ORAM is an iterated version of \cite{C:DemPapPap18} which uses this technique for their deamortization.
We believe that $\localORAM$ can be deamortized in the same manner but leave the details for future work.
\end{remark}

\begin{algorithm}[h]
\begin{flushleft}
\caption{Local Oblivious RAM ($\localORAM$)}
\label{algo:local_oram}
\underline{$\localORAM.\ORAMInit(1^\secpar, \Mem)$}
\begin{algorithmic}[1]
\State Parse $\Mem$ as $\{i,v_i\}_{i=1}^n$, where $\abs{i,v_i} = \orambs$
\State Let $n_1 = n^{\frac{1}{c}}$ and $n_i = n^{\frac{i}{c}} + n^{\frac{i-1}{c}}$ for $i\in[2,c]$
\State Let $A_i$ be an empty array of size $n_i$ for $i\in[1,c]$
\State Let $\pi_i : [1,n_i] \mapsto [1,n_i]$ be pseudorandom permutation for $i\in[2,c]$
\ForAll{$i\in[1,n]$} 
\State Store $(i,v_i)$ at locations $\pi_c[i]$ in $A_c$
\EndFor
\State Encrypt $A_i\encr \gets \Enc_{\KENC}(A_i)$ for $i\in[1,c]$ 
\State Let $R_i$ be an empty set (of maximal size $n^{i/c}$ blocks for $i\in[2,c-1]$) and $R_c = \Mem$ 
\State Set $\cnt_i \gets 0$ for $i \in [2,c]$ 
\State Let $\T_i$ be an empty hash table of size $n$ for $i\in[2,c-1]$ 
\State Encrypt $R_i\encr \gets \Enc_{\KENC}(R_i)$ and $\T_i\encr \gets \Enc_{\KENC}(\T_i)$ for $i\in[2,c-1]$  
\State Set $\st = (\{\pi_i\}_{i=2}^c, \{\cnt_i\}_{i=2}^{c}, \KENC)$
\State Set $\EMem = (\{A_i\encr\}_{i=1}^{c},\{R_i\encr\}_{i=2}^{c-1}, \{\T_i\encr\}_{i=2}^{c-1})$
\State \Return $\st, \EMem$
\end{algorithmic}
\underline{$\localORAM.\ORAMAccess(\st, k; \EMem)$}\\ 
\textit{Client:}
\begin{algorithmic}[1]
\State Retreive $(A_1\encr, \{\T_i\encr\}_{i=2}^{c-1})$ from the server and decrypt to $(A_1, \{\T_i\}_{i=2}^{c-1})$
\State Set $\found \gets false$ and $\cnt_i \gets \cnt_i + 1$ for $i\in[2,c]$
\If{$(k,v_k) \in A_1$}
\State $\found \gets true$
\EndIf
\ForAll{$i\in[2,c-1]$}
\If{$\found$ or $T_i[k] = 0$}
\State $\indA_i \gets \pi_i(n^{i/c} + \cnt_i)$ 
\Else
\State $\indA_i \gets \pi_i(T_i[k])$ 
\State $\found = true$
\EndIf
\State $\T_i[k]\gets \cnt_{i+1}$ 
\State $r_i \gets \Enc_{\KENC}(\cnt_{i+1},v_k)$ 
\EndFor
\If{$\found$}
\State $\indA_c \gets \pi_c[n + \cnt_c]$
\Else
\State $\indA_c \gets \pi_c[k]$
\EndIf
\State $A_1[\cnt_2] \gets (k,v_k)$
\State \Send $\{\indA_i\}_{i=2}^c$ and $\{r_i\}_{i=2}^{c-1}$
\end{algorithmic}
\textit{Server:}
\begin{algorithmic}[1]
\State Set $R_i \gets R_i \cup r_i$ for $i\in[2,c-1]$
\State \Send $\{A_i\encr[\indA_i]\}_{i=2}^c$ 
\end{algorithmic}
\textit{Client:}
\begin{algorithmic}[1]
\State Retreive block $(k,v_k)$ from either $A_1$ or (decrypted) $A_i[\indA_i]$ for some $i\in[2,c]$
\State Choose $i\in(c,\dots,2)$ maximal such that $\cnt_i > n^{\frac{i-1}{c}}$
\If{$i$ exists} 
\State Let $\pi_j$ be a new pseudorandom permutation for $j\in[1,i]$
\State Set $\cnt_j \gets 0$ for $j\in[2,i]$
\State Server updates $A_i$ with the result of $\OblSort(\pi_i,n_i,n^{1/c}\log^2 n; R_i)$
\State Empty $A_1$ and $\T_j$, such as $A_j$ and $R_j$ on the server, for $j\in[2,i-1]$
\EndIf
\State Store updated client state 
\State \Send reencrypted $(A_1\encr, \{\T_i\encr\}_{i=2}^{c-1})$ to server
\end{algorithmic}
\textit{Server:}
\begin{algorithmic}[1]
\State Update the encrypted memory $\EMem$ accordingly
\end{algorithmic}
\end{flushleft}
\end{algorithm}

\subsection{Tethys without Stash.}
\label{sub:stashless}
Now, we introduce a page-length-hiding static SSE scheme $\tethysoram$ that has $\bigO{\log^\eps \secpar}$ page efficiency, constant storage efficiency and contant client storage. 
We will later use it in the local transformation.

\subsubsection{More Preliminaries.}
Again, we require some additional preliminaries.

\begin{definition}[Binpacking]
We define the algorithm $\Binpack$ that takes at most $N$ keyword-identifier pairs $\Stash$ and a size $p$ as input. 
$\Binpack$ proceeds as follows. 
Allocate bins $B_1,\dots,B_{2N/p}$ and table $\TPos$.
Then, take list $L$ of identifiers matching keyword $w$ and insert the identifiers one-by-one into the bin with the smallest index that is not full yet.
Set $\TPos[w] = i$, where $i$ is the smallest index of a bin containing an identifier matching $w$.
Finally, fill the bins up to size $p$ with zeros.
Finally, output $(\Mem,\TPos)$.
\end{definition}
Clearly, if there are at most $p$ identifiers matching keyword $w$, the identifiers will all fit into $B_i,B_{i+1}$ for $i\gets \TPos[w]$. Also, note that $\Binpack$ can always fit all $N$ identifiers into the bins. 

\begin{lemma}[$\Tethys$ \cite{C:BBFMR21}]
\label{lemma:tethys}
The SSE scheme $\Tethys$ is correct and $\LH$-adaptively secure in the random oracle model (under the assumption that there exists an $\INDCPA$ secure encryption scheme and a secure pseudo-random function).
It has a client storage $\omega(\log \secpar)/\log N$ pages, and $\bigO{1}$ storage and page efficiency.
\end{lemma}
We call the client storage of $\Tethys$ its stash. 
In this work, a stash size of $\bigO{\log^{1+\delta}(\secpar)}=\omega(\log \secpar)/\log N)$ pages is sufficient for our construction, for some arbitrary $\delta > 0$.

\subsubsection{The Scheme.}
We now define the static SSE scheme $\tethysoram$ with client storage $\bigO{1}$. 
Let $p = \Omega(\secpar), c\in \N$ and $\delta > 1$.    
Essentially, we use $\Tethys$ (see \Cref{lemma:tethys}) and outsource its stash using $\localORAM$.
We define the static SSE scheme $\tethysoram$ for given page size $p$ as follows:

\para{$\tethysoram.\KeyGen(1^\secpar)$} 
Simply output $\K \gets \Tethys.\KeyGen(1^\secpar)$.

\para{$\tethysoram.\Setup(\K, N, \DB)$} 
The client generates encrypted database $\EDB'$ and stash $\Stash$ using $(\EDB',\Stash) \gets \Tethys.\Setup(\K,N,\DB)$. 
The stash contains the remaining keyword-identifier pairs $(w_i,\ind_i)$ that could not be allocated directly in $\Tethys$.
We want to outsource $\Stash$ using $\localORAM$ with $c$ levels and blocksize $\orambs = p$. The items in $\Stash$ are not necessarily lists of size $p$. Thus, we group $\Stash$ into pages of $p$ identifiers using $(\Mem,\TPos) \gets \Binpack(\Stash,p)$. Let $n = \bigO{\log^{1+\delta}(\secpar)}$.
After the binpacking, $\Mem$ consists of $n$ pages with $p$ identifiers each (see \Cref{lemma:tethys}) and we can access the identifiers matching keyword $w$ in page $i = \TPos[w]$, as there is at most one such page\footnote{Without loss of generality, we can assume that there are at most $p$ identifiers per keyword in the stash. For this, we can keep full lists inside a table $\TFull$ such as in $\pageSSE$. This version of $\Tethys$ was already described in \cite{C:BBFMR21} (see their scheme $\textsf{Pluto}$). Note further that the binpacking algorithm $\Binpack$ packs a list of identifiers into at most two consecutive bins. Thus, knowledge of $i$ sufficies to fetch bin $i$ and $i+1$. We assume in the following that the list of identifiers is in at most one bin for simplicity.}.
As this binpacking process is not data-oblivious, we can not leak $i$ to the server.
Thus, the client sets $\TPos\encr \gets \Enc_{\KENC}(\TPos)$ (after padding $\TPos$ to $n$ entries of $\log n$ bits).
Further, she applies $(\EMem, \st) \gets \localORAM.\ORAMInit(1^\secpar,\Mem)$.
Finally, she outputs $\EDB = (\EDB',\EMem,\TPos\encr)$ and stores state $\st$ locally.

\para{$\tethysoram.\Search(\K, w;\EDB)$}
The client initiates protocol $\Tethys.\Search(\K\allowbreak,w;\EDB)$ with the server from which she receives some of the identifiers matching keyword $w$. 
Next, the client retrieves $i\encr = \TPos\encr[w]$ from the server and decrypts the index of the page containing the remaining identifiers via $i\gets \Dec_{\KENC}(i\encr)$.
The client retrieves this page via $\localORAM.\ORAMAccess(\st, i; \EMem)$.

\begin{lemma}[$\tethysoram$]
\label{lemma:tethysoram}
The SSE scheme $\tethysoram$ is correct and $\LH$-adaptively secure. 
Let $\eps > 0$ and $p=\Omega(\secpar)$.
There are constants $\delta>1$ and $c\in \N$ such that it has constant client storage, constant storage efficiency and $\bigO{\log^\eps \secpar}$ page efficiency.
Further, the scheme has $\bigO{1}$ locality if each list fits into a constant number of pages.
\end{lemma}
\begin{proof}
We first show that $\tethysoram$ is correct and $\LH$-adaptively secure. 
Then, we analyze the efficiency for arbitrary constants $c$ and $\delta$ from which we conclude the existence of $\eps$.

(1) As $\Tethys$ is correct, it remains to show that all remaining identifiers are fetched from the stash when searching a keyword $w$.
First, note that we store $n = \bigO{\log^{1+\delta}(\secpar)}$ blocks in $\localORAM$. The scheme $\localORAM$ is correct if the blocksize $p$ is $\Omega{\left(n^{\frac{c-1}{c}}\right)} = \Omega{\left(\log^{\frac{1+\delta(c-1)}{c}}\secpar\right)}$ which holds as $\delta,c$ are constant and $p = \Omega(\secpar)$ by assumption.
As $\Binpack$ packs the stash into bins of size $p$ and the accessed index $i=\TPos[w]$ corresponds to the bin containing the identifiers matching keyword $w$, the scheme $\tethysoram$ is correct.

(2) The security follows directly as $\Tethys$ is $\L$-adaptively secure, $\localORAM$ is adaptively secure (with zero-failure probability) and the fact that $\TPos$ is encrypted.

(3) 
We now analyze the efficiency of $\tethysoram$.
As the client state of $\localORAM$ is $\bigO{1}$ and the instantiation of $\Tethys$ only stores its keys on the client side (as the stash is stored on the server), $\tethysoram$ only requires constant client storage.
The storage efficiency of $\tethysoram$ is $\bigO{1}$ because $\Tethys$ has constant storage efficiency, and $\EMem$ and $\TPos$ have size $n = \bigO{\log^{1+\delta}(\secpar)}$ pages and entries respectively.
We now inspect the page efficiency. 
First, note that $\Tethys$ has constant page efficiency and the access to $\TPos$ requires (at most) one page access. 
The access of the stash through $\localORAM$ requires bandwith $\bigO{\orambs \cdot n^{1/c}\log^2(n)} = \bigO{p \cdot \log^{\frac{1+\delta}{c}+\delta'}\secpar}$ for some arbitrary $\delta'>0$. As $c, \delta$ and $\delta'$ are arbitrary constants, scaling them accordingly yields the desired result. 
Constant locality follows directly from the fact that $\localORAM$ has constant locality and that $\Tethys$ accesses at most $\bigO{\ell/p}$ pages for a search on keyword $w$, where $\ell$ is the length of the list of identifiers matching $w$. (Recall that we assume that $\bigO{\ell/p}$ is constant for all lists.)
\end{proof}

\subsection{The Scheme}
\label{sub:uncond_scheme}
Finally, we describe our unconditional static SSE scheme $\uselessSSE$ with $\bigO{\log^\eps(N)}$ locality, for abitrary $\eps>0$. 
We follow the high level idea of \cite{C:DemPapPap18} to handle lists with different schemes depending on the list size.
For $d\in\N$, we split the interval $[1,N]$ of possible list lengths into four different subintervals.
\begin{enumerate}
    \item For the subinterval $[1,N^{1-1/\log\log \secpar})$, the lengths are sufficiently for $\pageSSE$ and can simply store the lists using $\local{\pageSSE}$. 
    Here, the read efficiency is $\bigO{\log\log N}$.
    \item For the subinterval $[N^{1-1/\log\log \secpar}, N/\log^d N)$, the lengths are simultaneously small enough for the local transformation and large enough for $\tethysoram$. Thus, we store the lists using $\local{\tethysoram}$ with $\bigO{\log^\eps N}$ read efficiency.
    \item We further split the subinterval $[N/\log^d N, N/\log^\eps N)$ into a constant number of subintervals, such that $\tethysoram$ has $\bigO{\log^\eps N}$ read efficiency.
    \item For the subinterval $[N/\log^\eps N, N]$, lists are large enough to read the entire database. Thus, we simply encrypt $\DB$ and fetch it entirely from the server for these lists.
\end{enumerate}

We now present how to divide the interval $[N/\log^d N, N/\log^\eps N)$ into subintervals in more detail.
\subsubsection{Handling the Remaining List Sizes.}
Note that for lists of size in $[N/\log^{\eps} N,N)$, we can just store an encrypted copy of the database on the server and retrieve the entire copy for each read.
We now sketch how we handle the remaining lists of size in $S = [N/\log^d N,N/\log^{\eps} N)$ for some arbitrary $d\in\N$ and $\eps \in \RIext{0}{1}$.
For this, we split the interval $S$ into a constant number of subintervals $S_i$ such that the borders of each interval differ by a factor $\log^{\eps}N$. Concretely, we set 
\[
    S_i = [N/\log^{d-i\eps} N,N/\log^{d-(i+1)\eps} N) \text{ for } i\in[0,\ceil{d/\eps}].
\]
For each $S_i$, we store lists of size in $S_i$ via $\tethysoram$ with page size $p = \max(S_i)$.
Note that each list has at most size $p$.
Thus, $\tethysoram$ has $\bigO{1}$ locality and $\bigO{\log^\eps \secpar}$ read efficiency (see \Cref{lemma:tethysoram}). 
Note that page efficiency directly translates to read efficiency in this case, as each list is at most a factor of $\log^\eps \secpar$ smaller than the page size.

\subsubsection{$\uselessSSE$.}
We now present our static SSE scheme $\uselessSSE$. For a given $\eps>0$, it has unconditionally $\bigO{\log^\eps N}$ read efficiency, constant locality and constant storage efficiency. Let $d\in\N$ be the parameter of the local transformation chosen appropriately.

\para{$\uselessSSE.\KeyGen(1^\secpar)$}
Generate key $\K_1$ for $\local{\pageSSE}$, key $\K_2$ for $\local{\tethysoram}$, key $\K_3$ for $\tethysoram$ and encryption key $\K_4$ for $\Enc$. Also, generate key $\KPRF$ for pseudorandom function $\PRF$ mapping to $\bit^{\ceil{\log(N)}}$. Output $\K = (\K_1, \K_2, K_3, K_4, \KPRF)$. 

\para{$\uselessSSE.\Setup(\K, N, \DB)$}
First, we initialize a table $\TLen$ that stores the encrypted length $\ell_i\xor m_i$ of each list $\DB(w_i)$ at position $\TLen[w]$, where $m_i \gets \PRF_{\KPRF}(w_i)$ is a mask. 
Then, we pad $\TLen$ up to size $N$ (with random values of $\ceil{\log N}$ bits).
We split the interval of possible list lengths $[1,N]$ into four different subintervals and handle each subinterval seperately.
For each subinterval, we define four databases $\DB_i$ containing a subset of the keyword-identifier pairs of the given database $\DB$ (chosen with respect to the lists length).
We set
\begin{align*}
    \DB_1 &= \left\{\DB(w_i) \mid \ell_i \in \left[1,N^{1-\frac{1}{\log\log N}}\right)\right\}, \\ 
    \DB_2 &= \left\{\DB(w_i) \mid \ell_i \in \left[N^{1-\frac{1}{\log\log N}}, N/\log^d N\right)\right\}, \\
    \DB_3 &= \left\{\DB(w_i) \mid \ell_i \in \left[N/\log^d N, N/\log^\eps N\right)\right\}, \\ 
    \DB_4 &= \left\{\DB(w_i) \mid \ell_i \in \left[N/\log^\eps N, N\right]\right\}.
\end{align*}
The lists in $\DB_1$ are sufficiently small for $\pageSSE$ and thus, we can apply the local transformation and run $\EDB_1 \gets \local{\pageSSE}.\Setup(\K_1,N,\DB_1)$. 
Note that we still pad the encrypted database to size $\bigO{N}$ and not $\bigO{\abs{\DB_1}}$ because we can not reveal the distribution of lists amongst each subinterval.
The lists in $\DB_2$ are sufficiently large for $\tethysoram$. 
Consequently, we can set $\EDB_2 \gets \local{\tethysoram}.\Setup(\K_2,N,\DB_2)$.
For $\DB_3$, we further split the interval $[N/\log^d N, N/\log^\eps N]$ into the constant number of subintervals $S_i = \RI{N/\log^{d-i\eps} N}{N/\log^{d-(i+1)\eps} N}$ for $i\in[0,\ceil{d\cdot \eps^{-1}}]$ as described above. 
We then set $\DB_{3,i} = \left\{\DB(w_i) \mid \ell_i \in S_i \right\}$ and $\EDB_{3,i} \gets \tethysoram.\Setup(\K_3,N,\DB_{3,i})$. 
Finally, set $\EDB_3 = (\EDB_{3,1}, \cdots, \EDB_{3,\ceil{d\cdot \eps^{-1}}})$.
Lastly, lists in $\DB_4$ are large enough that we can scan entire database each read. 
For this, we pad $\DB_4$ up to size $N$ and set $\EDB_4 \gets \Enc_{\KENC}(\DB_4)$.
Outputs $\EDB = (\EDB_1,\EDB_2,\EDB_3,\EDB_4,\allowbreak \TLen)$.

\para{$\uselessSSE.\Search(\K, w;\EDB)$}
For retreiving the identifiers matching keyword $w$, the client sends $w$ and $m \gets \PRF_{\KPRF}(w)$ to the server.
The server decrypts the length $\ell \gets \TLen[w] \xor m_i$ of the list to be fetched and then checks in which subinterval $\ell$ lies. We distinguish four cases:
(1) If $\ell \in [1,N^{1-\frac{1}{\log\log N}})$, the client retrieves the identifiers from the server via $\local{\pageSSE}.\Search(\K_1, w;\EDB_1)$.
(2) If $\ell \in [N^{1-\frac{1}{\log\log N}}, N/\log^d N)$, the client runs $\local{\tethysoram}.\Search(\K_2, w;\EDB_2)$ with the server.
(3) If $\ell \in [N/\log^d N, N/\log^\eps N)$, the server sets $i\in[0,\ceil{d\cdot \eps^{-1}}]$ such that $\ell\in S_i$. 
Then, server and client run $\tethysoram.\Search(\K_3, w;\EDB_{3,i})$.
(4) Otherwise, we have $\ell \geq N/\log^\eps N$ and the server sends the entire encrypted database $\EDB_4$ to the client (from which he fetches the corresponding list).

\begin{theorem}[$\uselessSSE$]
The scheme $\uselessSSE$ is correct and $\LR$-adaptively secure.
It has constant client storage, $\bigO{1}$ locality and $\bigO{\log^\eps N}$ read efficiency for any $\eps > 0$. 
\end{theorem}
\begin{proof}
Security (and correctness) directly follow from the security of $\local{\pageSSE}$, $\local{\tethysoram}$ and $\tethysoram$.
The efficiency properties of $\uselessSSE$ can also be derived from the efficiency properties of the used SSE schemes (see discussion above).
\end{proof}

\begin{remark}[On RTT and Deamortization]
The scheme $\uselessSSE$ uses $\localORAM$ and thus, the efficiency properties are amortized. 
Also, this introduces a large round trip time for some operations.
We note that if $\localORAM$ is deamortized, we can adapt $\uselessSSE$ in order to have a constant RTT and deamortized efficiency.
\end{remark}
\section{Analysis of $\layeredchoice$}
\label{sec:alloc_proof}
Here, we proof \Cref{thm:layeredchoice}.
First, we introduce some additional preliminaries.

\begin{lemma}[Chernoff’s Bound]
\label{lem:chernoff}
Suppose that $X_1,\dots,X_n$ are independent random variables taking values in $\{0,1\}$. Let $X$ denote their sum and let $\mu = \Exp[X]$ denote the expectancy of $X$. Then for any $\delta > 0$, it holds that
\[
\Pr[X < (1-\delta)\mu] \leq e^{-\frac{\delta^2\mu}{2}}
\]
\end{lemma}

In the next lemma, we consider a sequence of ball insertions and deletions of arbitrary length, such that the total number of balls in the bins at any point in time is bounded by $n = h\cdot m$. A ball insertion is a standard 2-choice insertion: pick two bins i.u.r., and insert the ball into the least loaded bin. A deletion removes one previously inserted ball. The sequence of additions and deletions is fixed at the input of the problem.

\begin{lemma}[$\twochoice$]
\label{lemma:twochoice}
Let $\delta(m)$ be an arbitrary map such that $1 \leq \delta(m) \leq \log m$ for all $m \geq 1$.
At the outcome of the sequence of additions and deletions, the most loaded bin contains $O(h + \delta(m) \log \log m)$ items, except with probability $m^{-\Omega(\delta(m) \log \log m)}$.

In particular,
by setting $\delta = 1$, we get that if $m \geq \lambda$, then the failure probability from the claim is negligible.
By setting $\delta = \log\log\log m$, we get that if $m \geq \lambda^{1/\log \log \lambda}$, then the failure probability from the claim is negligible.
\end{lemma}

\begin{proof}
We adapt the proof of \cite{vocking2003asymmetry}, which proves a bound $O(h) + \log \log m$ with probability $m^{-\alpha}$, for an arbitrary constant $\alpha$.
The proof uses \emph{witness trees}. The existence of a bin containing more than $Ch + L$ items implies the existence of a witness tree of height $L + C'$, for some suitable constants $C$, $C'$. Thus, in order to bound the probability that a bin contains more than $Ch+L$ items, it suffices to bound the probability that a witness tree of height $L+C'$ exists. In more detail, the proof shows that the probability that a witness tree of height $L+3$ exists is upper-bounded by
\[
m^{-\kappa+1+o(1)} + m^{-\alpha}
\]
where $\kappa$, $\alpha$ are certain parameters (to be discussed later), with:
\[
L \leq \log \log m + \log(1+\alpha) + \kappa.
\]

The proof sets $\alpha$ and $\kappa$ to be constants.
The fact that $\gamma$ and $\kappa$ are constant is not essential to the argument, and is only used in two places in the proof.

The first place is the end of Section~2.3, when upper-bounding the probability of activation of a pruned witness tree by $m^{-\kappa+1+o(1)}$. The final step of that upper-bound requires $\alpha\cdot\kappa = m^{o(1)}$, which is obviously true for a pair of constants.

The other, more important place where the choice of having constant $\alpha$ and $\kappa$ comes into play is in the final derivation.
The proof shows that, except with probability at most $m^{-\kappa+1+o(1)} + m^{-\alpha}$, the number of items in the most loaded bin is at most:
\begin{align*}
L + O(h) &\leq \log \log m + \log(1+\alpha) + \kappa + O(h)\\
&= \log \log m + O(1) + O(h)\\
&=\log \log m + O(h).
\end{align*}
In that final computation, the fact that $\alpha$ and $\kappa$ are constant makes it possible to absorb the $\log(1+\alpha) + \kappa$ term into the $O(h)$ term. The other term is only $\log \log m$, which is optimal.
If we set $\alpha = \kappa = \delta(m) \log \log m$  instead, we get:
\begin{align*}
L + O(h) &\leq \log \log m + \log(1+\alpha) + \kappa + O(h)\\
&\leq 3\delta(m)\log \log m + O(h).
\end{align*}
In the case $\delta = 1$, this worsens the constant in front of the $\log \log$ term, which is likely why the authors chose $\alpha$ and $\kappa$ to be constant.
(A better constant than 3 is possible, we choose 3 for simplicity.)
On the other hand, the probability of failure becomes at most
\[
m^{-\kappa+1+o(1)} + m^{-\alpha} = m^{-\Omega(\delta(m)\log \log m)}
\]
as claimed. Note that the condition $\alpha\cdot\kappa = m^{o(1)}$ is still fulfilled.
\end{proof}

The next lemma is a direct application of Markov's inequality.

\begin{lemma}
\label{lem:firstmoment}
For any random variable $X \in [0,N]_\R$ and any $R > 0$ (which may depend on $N$):
\[
\Pr{[X > R]} = \negl\quad \text{iff}\quad \Exp[{\max(X - R,0)}] = \negl.
\]
\end{lemma}

\begin{lemma}[Weigthed $\onechoice$ \cite{berenbrink2008weighted}.]
\label{lem:weightedballs}
Let $m\in[0,1]_\R$ be some maximal weight.
Let $x = (m)_{i\leq n}$ and $x' = (w_i')_{i\leq n'}$ be (non-negative) weight vectors. 
Let $\sum_{i=1}^{n'} w_i' \leq n\cdot m$ and $w_i \leq m$ for all $i\in\Int{n'}$\footnote{\cite{berenbrink2008weighted} requires that $x$ majorizes $x'$. This is implied by our condition on $x$ and $x'$.}.
Let $R \in \R^+$. 
Then it holds that 
$\Exp[\max(X_\mathsf{mlb}-R, 0)] \geq \Exp[\max(X_\mathsf{mlb}'-R, 0)]$, where $X_\mathsf{mlb}$ ($X_\mathsf{mlb}'$) are a random variable indicating the load of the most loaded bin after throwing $n$ balls with weights $x$ ($n'$ balls with weights $x'$) uniformly and independently at random into $m$ bins\footnote{\cite{berenbrink2008weighted} shows that $\Exp[X_\mathsf{mlb}] \geq \Exp[X_\mathsf{mlb}']$. As $f(X) = \max(X-R, 0)$ is convex, their proof can be adapted to our formulization.}. 

In words, for $\onechoice$, the load above threshold $R$ of the most loaded bin is higher with balls of weight $x$ than with balls of weight $x'$.
\end{lemma}
    
We are now ready to prove \Cref{thm:layeredchoice}.
\begin{proof}
Note that the load of a bin is never decreasing, so it is sufficient to analyze the final load of bins $B_1,\dots,B_m$. 
Also, note that we can replace $\Setup$ with $n$ $\InsertBall$ operations. 
Thus, we can assume without loss of generality that bins $B_1,\dots B_m$ are initially empty after $\layeredchoice.\Setup$.
Also, note that $m^{-\Omega(\delta(\secpar)\log\log w)}=\negl$ under the given requirements (see \cref{lemma:twochoice}).
As $\Hash$ is modeled as a random oralce, we assume that the bin choices $\alpha_1,\alpha_2$ of ball $b$ are chosen independently and uniformly at random from $[1,m]^2$.
We split the proof into three parts:

(1) First, we will modify the sequence $S$ such that we can reduce the analysis to only (sufficiently independent) $\layeredchoice.\InsertBall$ operations, while only increasing the final bin load by a constant factor.

(2) Second, we analyze the maximal bin load when only considering balls of weight at most $1/\log m$. 
Here, $\layeredchoice.\InsertBall$ proceeds exactly as weighted $\onechoice$. 
Since uniform weights of value $1/\log m$ are the worst case for the most loaded bin in $\onechoice$, the bound follows from a Chernoff’s bound as balls are sufficiently small.

(3) Last, we inspect the maximal bin load considering items in the remaining subintervalls
$(2^{i -1}/\log m, 2^i/\log m]_\R$ for $i\in\Int{\log\log m}$.  
Per interval, $\layeredchoice.\InsertBall$ behaves like unweighted two-choice (independent of other subintervals) and inherits the $\log \log m$ bin load direclty, as balls with different weights differ only by a constant factor per interval.
Summing up the maximal bin load per interval will yield the desired result. 

\textbf{Part 1 -- Adapting the sequence:}
We observe that update operations updating the weight inside the same subinterval can be ignored.
More concretely, let $\op_i = \UpdateBall$ be some update operation on ball $b_i$ with old weight $o_i$ and new weight $w_i$.
If $o_i, w_i \in (\frac{2^{k-1}}{\log m},\frac{2^{k}}{\log m}]_\R$ for some $k$, the operation $\op_i$ replaces the old weight $o_i$ of ball $b_i$ with the new weight $w_i$ direclty (inside the same bin).
Thus, we can simply remove $\op_i$ and replace the previous operation $\op_j = (b_i, o_j, o_i) = (b_j, o_j, w_j)$ on the same ball with $\op_j' = (b_i, o_j, w_i)$ directly.
Clearly this does not change the final load of the bins.
(Note that operations between $\op_j$ and $\op_i$ make the same choices as the concrete weight inside a subinterval never impacts which bin is chosen.)

Now, let $(\op_i)_{i\in I}$ be all remaining update operations for some fixed ball $b_*$, so $b_* = b_i$ and $\op_i = \UpdateBall$ for $i\in I$.
As we removed consecutive update operations in the same subinterval, operation $\op_i$ marks the ball $b_*$ as \leftover ball and calls $\InsertBall(b_*,w_i,B_{\alpha_{*,1}},B_{\alpha_{*,2}})$.
Let $j = \max(I)$ be the index of the last update operation $\op_j$ on $b_*$ and $k$ be minimal such that $w_j \leq 2^k/\log m$. 
As there are only $k$ subintervals below the last interval $(2^{k-1}/\log m, 2^k/\log(m)]_\R$, there are at most $k$ such update operations, \ie $\abs{I} \leq k$, and one insert ball operation.
Assume without loss of generality that all $k+1$ operations exist. 
The \leftover ball left by the $i$-th update operation has at most size $2^{i-1}/\log m$ and thus, this $\UpdateBall$ operation can be replaced by an $\InsertBall(b_*,2^{i-1}/\log m, (B_{\alpha_{*,1}},B_{\alpha_{*,2}})$ operation.
Thus, for ball $b_*$ with final weight $w_j$, we have to insert $k$ additional balls in order to replace all update operations on ball $b_*$ with inserts. The total weight of these additional balls is 
\[
    \sum_{i=1}^k 2^{i-1}/\log m \leq 2^k/\log(m) \leq 2w_j,
\]
since $w_j \geq 2^{k-1}/\log(m)$. Thus, the total weight is increased at most by a factor $3$ per ball.

This way, we can iteratively remove all $\UpdateBall$ operations at the cost of a factor $3$ in the total weight. 
The remaining operations are $\InsertBall$ operations, where each ball $b_i$ is inserted at most once per subinterval and the bin choices are drawn uniformly and independently random per ball.
Clearly, if $\bigO{3\log\log \maxweight}$ is an upper bound on the load of the most loaded bin for the modified sequence $S'$, then $\bigO{3\log\log \maxweight}$ is an upper bound for the initial sequence $S$. 
In the following, we only consider modified sequences $S$ of $n$ such $\InsertBall$ operations.

\textbf{Part 2 -- Light balls:}
Here, we show that the most loaded bin has load at most $3\delta(\secpar)\log \log \maxweight$ when only considering balls of at most weight $1/\log m$. 
Let $w\leq \maxweight$ be the total weight of all such light balls. 
Without loss of generality, assume that $\maxweight = w$.
At first, we assume that all such balls have weight exactly $1/\log m$ each.
We will then reduce the case with arbitrary weights in $[0,1/\log m]_\R$ to the above.

Since we initially assume all balls have weight $1/\log m$, the number of balls is at most $n' = w\log m$.
Let $X_i$ be the random variable that denotes the number of balls in bin $B_i$. 
Recall that $m = \frac{w}{\delta(\secpar)\log\log w}$. 
We observe that $\InsertBall$ behaves like $\onechoice$ in this case and thus, we have $\Exp[X_i] = n'/m = \delta(\secpar)\log\log w \cdot \log m$.
Applying Chernoff’s bound (\Cref{lem:chernoff}), we get:
\[
    \Pr[X_i \geq (1+\gamma)\Exp[X_i]] \leq \exp\left({-\frac{\gamma^2\Exp[X_i]}{2}}\right).
\]
We insert $\gamma = 2$ in the equation above and receive:
\[
    \Pr[X_i \geq 3\delta(\secpar)\log\log w \cdot \log m] \leq \exp(-2\delta(\secpar)\log\log w \cdot \log m) 
                                              = m^{-\Omega(\delta(\secpar)\log\log w)}.
\]
A union bound yields that the most loaded bin contains at most $3\delta(\secpar)\log\log w \cdot \log m$ balls with probability at most $m^{-\Omega(\delta(\secpar)\log\log w)} = \negl$. 
As each ball has size $1/\log m$, the most loaded bin has a maximal load of $3\delta(\secpar)\log\log w$ with overwhelming probability.

Now, we show this bound is preserved when allowing arbitrary weights of at most $1/\log m$.
We define the weight vectors $x = (1/\log m)_{i=1}^{w\log m}$ and $ x' = (w_i)_{i\in I}$. 
Let $X_\mathsf{mlb}$ and $X_\mathsf{mlb}'$ be the random variable indicating the load of the most loaded bin with weights $x$ and with weights $x'$ respectively. We want to show that $\Pr[X_\mathsf{mlb}' > 3\delta(\secpar)\log\log w] = \negl$.
\Cref{lem:firstmoment} implies that the above holds iff $\Exp[\max(X_\mathsf{mlb}' -3\delta(\secpar)\log\log w), 0)] = \negl$. 
This expectancy can be upper bound by $\Exp[\max(X_\mathsf{mlb} -3\delta(\secpar)\log\log w), 0)]$ as $x$ and $x'$ fulfill the requirements of \Cref{lem:weightedballs}.
As we showed above that $\Pr[X_\mathsf{mlb} > 3\delta(\secpar)\log\log w)] = \negl$, we can conclude from another application of \Cref{lem:firstmoment}. 

\textbf{Part 3 -- Heavy balls:}
So far, we have shown that the most loaded bin has load at most $3\delta(\secpar)\log\log w$ with overwhelming probability, when only considering balls of weight smaller or equal to $1/\log m$, for any (modified) sequence $S$.
We will now show that when considering the remaining balls of weight in $(1/\log m, 1]_\R$, a maximal load of $\bigO{\maxweight/m  + \delta(\secpar)\log\log w} = \bigO{\delta(\secpar)\log\log w}$ is preserved.

For $i\in[1,\log\log m]$, let $n_i$ be the number of balls in each subinterval $A_i = (2^{i -1}/\log m, 2^i/\log m]_\R$. 
Recall that each ball $b_i$ has two bin choices that are drawn uniformly and independently random at the first insertion. 
These choices are reutilized across the subintervals $A_i$, if $b_i$ is inserted in multiple subintervals. But note that per subinterval, $b_i$ is only inserted once.
Thus, $\layeredchoice$ behaves like unweighted $\twochoice$ on all balls with weights in $A_i$ (independent from the balls in other subintervals). 
By \Cref{lemma:twochoice}, the bin with the highest number of balls (of weights in $A_i$) contains at most $\bigO{n_i/m + \delta(\secpar)\log\log m}$ balls with overwhelming probability. (Note that there are at most $w\log m$ balls and that $m^{-\Omega(\delta(\secpar)\log\log w)} = \negl$.)

Each ball has weight at most $\max(A_i) = 2^i/\log m$ and thus, the load of the most loaded bin is at most $2^i/\log m(\bigO{n_i/m + \delta(\secpar)\log\log m})$ when considering balls with weights in $A_i$.
Summing over all $A_i$'s, when considering only balls with weights in $(1/\log m, 1]$, the load of the most loaded bin is at most
\begin{align*}
          &\sum_{i=1}^{\log\log m} {\frac{2^i}{\log m}\bigO{n_i/m + \delta(\secpar)\log\log m}}\\
     =    &\sum_{i=1}^{\log\log m} \bigO{2\frac{n_i2^{i-1}}{m\log m}
            + \sum_{i=1}^{\log\log m} \frac{2^i}{\log m} \bigO{\delta(\secpar)\log\log m}}\\
    \leq  &\bigO{\frac{\maxweight}{m} + \delta(\secpar)\log\log \maxweight},
\end{align*}
as $m=\bigO{\maxweight}$ and $\maxweight$ is an upper bound on the total weight.
The above holds with overwhelming probability, since the probability that $\twochoice$ fails is $\negl$, and there are only $\log\log m$ subintervals.

As we showed in the first part that is suffices to look at the modified sequence (with only $\InsertBall$ operations), we conclude that the load of the most loaded bin is at most $\bigO{\log\log \maxweight}$. 
\end{proof}


\section{Security Analysis of $\pageSSE$}
\label{sec:peff_sec}

\begin{lemma}[Correctness]
The scheme $\pageSSE$ is correct if at most $p$ identifiers are associated to each keyword and $\Hash$ is modeled as a random oracle.
\end{lemma}

\begin{proof}
We use $\layeredchoice$ to insert (and update) the lists of identifiers $\DB(w)$ of length $\ell\leq p$ into $m$ bins.
Each list is interpreted as a ball of weight $\ell/p \in [0,1]$.
\Cref{thm:layeredchoice} implies that the maximal loaded bin has load at most $c\log\log\log(\secpar)\allowbreak\log\log(N/p)$ for some appropriate constant $c\in \N$ (for $\delta(\secpar) = \log\log\log(\secpar)$), since the bin choices via $\hash$ are uniformly and independently random by assumption.
That means, it contains at most $p\cdot c\log\log\log(\secpar)\log\log(N/p)$ identifiers (as we scaled weights by a factor $p$).
Consequently, the bins only overflow with negligible probability.
Further, it follows from inspection that one of the two bins returned by the search algorithm on input $w$ contains all the identifiers matching keyword $w$.
\end{proof}

\begin{lemma}[Selective Security]
Let $\L_\Stp(\DB, N) = N$, $\L_\Srch(w) = \qpat$ and $\L_\Updt(\op,w,L') = \qpat$, where $\qpat$ is the query pattern and $\op = \add$.
Let $\L = (\L_\Stp, \L_\Srch, \L_\Updt)$. 
The scheme $\pageSSE$ is $\L$-selectively semantically secure if at most $p$ identifiers are associated to each keyword, $\Enc$ is $\INDCPA$ secure and $\Hash$ is modeled as a random oracle.
Note that $\L = \LH$ because we restrict ourselves to lists of size at most $p$.
\end{lemma}
\begin{proof}
Let $\Sim$ denote the simulator and $\adv$ an abitrary honest-but-curious PPT the adversary.

Initially, $\Sim$ receives $\L_\Stp(\DB,N) = N$ and a series of search and update requests with input $\L_\Updt(\op_i,w_i,L_i') = \L_\Srch(w_i) = \qpat$.
First, $\Sim$ initializes $m = \ceil{(N/p)/(\log\log (N/p)\log\log\log(\secpar))}$ bins $B_1,\dots,B_m$ zeroed out up to size $p\cdot c\log\log\log(\secpar)\log\log (N/p)$, and outputs $\EDB' = (\Enc_{\KENC'}(B_1), \dots, \Enc_{\KENC'}(B_m))$ for some encryption key $\KENC'$ sampled by $\Sim$.
Next, $\Sim$ simulates the search and update queries. 

For search queries, $\Sim$ receives $\spat$. 
If the query pattern $\spat$ indicates that the keyword was already queried, $\Sim$ outputs the keyword $w'$ from the previous query.
Otherwise, $\Sim$ outputs a new uniformly random keyword $w'$ (that has not been queried yet). 

For update queries, $\Sim$ receives $\spat$. First, $\Sim$ proceeds as in search for generating the first output $w'$.
After sending $w'$ to the adversary $\adv$, $\Sim$ receives two encrypted bins. 
$\Sim$ simply reencrypts both bins and sends them back to the server.

We now show that the real game is indistinguishable from the ideal game. For this, we define four hybrid games.
\begin{itemize}
    \item {Hybrid 0} is identical to the real game.
    \item {Hybrid 1} is the same as {Hybrid 0} except the simulated keywords $w'$ are output.
    By assumption Hybrid 0 and Hybrid 1 are indistinguishable.
    \item {Hybrid 2} is the same as Hybrid 1 except a flag $\FAIL$ is raised when a bin overflows (\ie contains more than $p\cdot c\log\log N/p$ identifiers after $\Setup$ or $\Update$).
    \Cref{thm:layeredchoice} implies that this happens only with negligible probability.
    Thus, Hybrid 1 and Hybrid 2 are indistinguishable.
    \item {Hybrid 3} is the same as Hybrid 2 except that the encrypted database $\EDB$ is replaced with the simulated $\EDB'$ and bins are just reencrypted and sent back to the adversary in the second flow of $\Update$. Since $\Enc$ is $\INDCPA$ secure (and a flag $\FAIL$ is only raised with negligible probability), it follows that Hybrid 2 and Hybrid 3 are indistinguishable.
    \item {Hybrid 4} is the same as the ideal experiment. The server's view in the ideal experiment and in Hybrid 3 are identically distributed, so we conclude inductively that the ideal game and the real game are indistinguishable.\qedhere
\end{itemize}
\end{proof}
\section{Proof Of $\clippedOSSE$}
\label{sec:clippedOSSE}

\newcommand{\E}[1]{\ensuremath{\mathbb{E}\left[#1\right]}\xspace}
\renewcommand{\Pr}[1]{\ensuremath{\mathsf{Pr}\left[#1\right]}\xspace}
\newcommand{\Bin}{\ensuremath{\mathsf{Bin}}}
\newcommand{\inddd}{\ensuremath{\mathsf{ind}}}
\newcommand{\cut}{\ensuremath{\mathsf{cut}}}
\newcommand{\ov}{\ensuremath{\mathsf{over}}}
\newcommand{\iter}{\ensuremath{\mathsf{merge}}}

In this section, we prove \Cref{thm:clippedOSSE}.
Recall that it is assumed all lists have length at most $N/ \log^d \lambda$ for $d \geq 2$. This is a limitation of the result, and it is inherent (lists of length close to $N/\log \lambda$ can create too many overflowing elements, and must be handled separately.)

The proof is divided into two parts. First, we show that the result holds when all lists have size exactly $N/\log^d \lambda$. Second, we show that the result still holds as long as all lists have size at most $N/\log^d \lambda$. The second part is the hard part.

\subsection{Proof Part 1: All Lists Have Size $N/\log^d \lambda$.}

We recall one of the standard formulations of the Chernoff-Hoeffding bound.

\begin{lemma}[Chernoff-Hoeffding]
\label{lem:chernoff}
Let $X = \sum_{i \leq n} X_i$ where the $X_i$'s are i.i.d. 0-1 random variables, with $p = \E{X_i}$.
\begin{align*}
\Pr{X > (p+\varepsilon)n} < e^{-D(p+\varepsilon \| p)n} &< e^{-\frac{1}{2}\varepsilon^2 n/(p+\varepsilon)}\\
\Pr{X < (p-\varepsilon)n} < e^{-D(p-\varepsilon \| p)n} &< e^{-\frac{1}{2}\varepsilon^2 n/p}.
\end{align*}
\end{lemma}

The following lemma is a direct corollary.

\begin{lemma}
Throw $n$ balls into $m$ bins u.i.r.
Let $\mu = n/m$ be the average load of a bin.
Then the probability that a given bin contains more than $\gamma \mu$ balls is at most:
\[
e^{-\Theta(\gamma)\mu}.
\]
\end{lemma}

\begin{proof}
Use Chernoff-Hoeffding with $p = 1/m$, $\varepsilon = (\gamma-1)p$.
\end{proof}

Let $\tau = \beta \log \log \lambda$ be the threshold at which a bin starts to overflow.

By construction, 1C with all lists of size $N/\log^d \lambda$ is exactly a balls-and-bins game with $n = \log^d \lambda$ balls (each ball is a list) and $m = n/\log \log \lambda$ bins.
Previous lemma says that the most loaded bin contains less than $\log \lambda$ elements, except with negligible probability.
We now want to bound the number of bins that overflow (= contain more  than $\tau$ elements).
Using the previous lemma again, the probability that a given bucket overflows is $e^{-\Omega(\beta) \log \log \lambda} = \log^{-\Omega(\beta)} \lambda$.

Let $(X_i)_{i \leq m}$ denote the indicator variables that are equal to 1 iff the $i$-th bucket overflows, 0 otherwise.
The number of overflowing buckets is $X = \sum X_i$.
We know that $\E{X_i} = \log^{-\Omega(\beta)} \lambda$.
We want to show that $X$ cannot be much higher than $m\log^{-\Omega(\beta)} \lambda$.

For that purpose, we use the notion of negative association. Because that notion is only used briefly to establish that Chernoff-Hoeffding bounds apply, we do not develop the theory here, and instead refer the reader to~\cite{negative} for an excellent survey on the topic.
By \cite[Proposition~13]{negative}, the occupancy numbers (vector $(B_{n,m}[i])_{i\leq m}$ where $B_{n,m}[i]$ is the number of balls in the $i$-th bin) are negatively associated. By \cite[Proposition~7.2]{negative} in the same reference, since $X_i = \mathbf{1}_{B_{n,m}[i] \geq \tau}$, and $x \mapsto \mathbf{1}_{x \geq \tau}$ is non-decreasing, the $X_i$'s are also negatively associated. By \cite[Proposition~5]{negative}, it follows that we can apply Chernoff-Hoeffding bounds to $X = \sum X_i$.

Hence, using \Cref{lem:chernoff} with $p = \varepsilon = \E{X_i} = \log^{-\Omega(\beta)} \lambda$, we get:
\[
\Pr{X > 2m\log^{-\Omega(\beta)} \lambda} = e^{-\frac{1}{4}\varepsilon m} = e^{-\frac{1}{4}m\log^{-\Omega(\beta)} \lambda}.
\]
The above quantity is negligible as soon as $d - \Omega(\beta) \geq 2$. (Recall that $d$ is such that the longest list is of size at most $N/\log^d N$.)
The above computation is loose and in the actual article, we can get concrete values for $\beta$ and $d$ if we want.

Since at most $2m\log^{-\Omega(\beta)} \lambda$ buckets overflow, and the most loaded bucket contains at most $\log \lambda$ items, we get that with overwhelming probability, the number of overflowing balls is less than:
\[
\log^d \lambda \log^{1-\Omega(\beta)} \lambda.
\]
Since each ball corresponds to a list containing $N/\log^d \lambda$ items, with overwhelming probability the number of overflowing \emph{items} is:
\[
N \log^{1-\Omega(\beta)} \lambda
\]
so we can make it $O(N/\log^c \lambda)$ for any constant $c$ of our choice by picking $\beta$ suitably (and then picking $d$ to satisfy the condition $d - \Omega(\beta) \geq 2$ encountered earlier).

\subsection{Proof Part 2: General Case}

Let $L$ denote an arbitrary multiset of list lengths, with $\max L \leq N/\log^d \lambda$, and $\sum L = N$. Let $X_L$ be (the random variable denoting) the number of overflowing elements after inserting the lists in $L$.

In Part 1 of the proof, we have seen that $\Pr{X_L > R} = \negl(\lambda)$, for a suitable $R$, when all lists are size $N/\log^d \lambda$. Our goal is to show a similar result for arbitrary $L$.


\subsubsection{Preliminaries}

We are going to work with $\E{\max(X_L - R,0)}$, instead of $\Pr{X_L > R}$. This is made possible by the following lemma.

\begin{lemma}
\label{lem:firstmoment}
For any random variable $X \in [0,N]$ and any $R \geq 0$:
\[
\Pr{X > R} = \negl(\lambda)\quad \text{iff}\quad \E{\max(X - R,0)} = \negl(\lambda).
\]
\end{lemma}

\begin{proof}
By a classic inequality, for any positive integral random variable $Y$, $\E{Y} = \sum_{i \geq 0} \Pr{Y > i}$. It follows that $\Pr{Y > 0} \leq \E{Y}$.
On the other hand, if $Y \leq N$, we get $\E{Y} \leq N\Pr{Y > 0}$. Hence:
\[
\Pr{Y > 0} \leq \E{Y} \leq N \Pr{Y > 0}.
\]
Since $N = \poly(\lambda)$, it follows that $\Pr{Y > 0}$ is negligible iff $\E{Y}$ is negligible. The lemma is obtained by applying that observation to $Y = \max(X - R,0)$.
\end{proof}


The next lemma is a corollary of Markov's inequality.

\begin{lemma}
\label{lem:condexp}
Let $X$, $Y$ be two random variables defined on the same sample space.
Let $\mathcal{E}$ be a set of events that forms a partition of the sample space (i.e. pairwise disjoint events whose union is the whole space).
If the conditional expectations satisfy $\E{X : E} \leq \E{Y : E}$ for all $E \in \mathcal{E}$, then $\E{X} \leq \E{Y}$.
\end{lemma}

\subsubsection{Notation}

\begin{itemize}
\item Let $B(p)$ denote the Bernoulli distribution with mean $p$: that is, a sample of $B(p)$ is a 0-1 random variable $X$ such that $\Pr{X = 1} = p$ and $\Pr{X = 0} = 1-p$.
\item Let $\Bin(p,n)$ denote the Binomial distribution with $n$ trials, each with probability $p$: that is, a sample of $\Bin(p,n)$ is distributed like $\sum_{i \leq n} X_i$, where the $X_i$'s are i.i.d. drawn from $B(p)$.
\item Suppose we throw $n$ balls i.u.r. into $m$ buckets. Let $B_{n,m}[i]$ be the (random variable denoting the) load of the $i$-th bucket. Let $B_{n,m} = (B_{n,m}[i])_{i \leq m}$ be the vector of the load of buckets. Observe that $B_{n,m}[i]$ is distributed according to $\Bin(1/m,n)$. Also note that the $B_{n,m}[i]$'s are not independent, e.g. they are linked by $\sum B_{n,m}[i] = n$.
\item If $D$ is a distribution, and $E$ is an event, then $D[E]$ denotes the distribution $D$ conditioned on the event $E$.
\item If $X$ is a random variable, and $E$ is an event, then $\E{X : E}$ denotes the conditional expectation of $X$, conditioned on the event $E$.
\item If $D$ is a distribution, $X_1, \dots, X_n \hookleftarrow D$ denotes that $X_1, \dots, X_n$ are i.i.d. random variables, each distributed according to $D$.
\end{itemize}

Recall that we are trying to show that when inserting lists according to algorithm 1C (specified at the start of this note), the number of overflowing items is bounded by some $R = O(N/\log^c N)$, except with negligible probability, for a constant $c$ of our choice. In Part~1 of the proof, we have already seen that this holds true when all lists are of size $N/\log^d N$, for some suitable constant $d$. In Part~2, we want to prove the same for an arbitrary multiset $L$ of list sizes, assuming $\max L \leq N/\log^d N$. Recall that $\sum L = N$.
Let $\tau = \beta \log \log \lambda$ be the threshold at which buckets are cut off.
For a given bucket load vector $b = (b[i])_{i \leq m}$, let $\ov(b)$ denote the number of overflowing items: $\ov(b) = \sum \max(b[i] - \tau,0)$.

Fix a multiset $L$ of list sizes, with $\max L \leq N/\log^d N$. Let $N_\ell$ denote the number of lists of size $\ell = 2^i$. Note $\sum \ell N_\ell = N$. Recall that the number of buckets is $m = N/\log \log \lambda$. (Relative to the notation earlier, we set $\alpha = 1$, because it is starting to look like $\alpha$ was useless.)

Let $D(L)$ be (the random variable denoting) the load of buckets at the output of algorithm 1C, on input $L$.
By abuse of notation, we still write $D(L)$ for a random variable distributed according to $D(L)$.
Our goal is to show $\Pr{\ov(D(L)) > R} = \negl(\lambda)$.
By Step~1, this is equivalent to $\E{\max(\ov(D(L)) - R,0))} = \negl(\lambda)$.
To simplify notation, write $F(L) = \max(\ov(D(L)) - R,0))$.

\subsubsection{Outline}


Starting from $L$, let $\mu = \min L$ be the smallest list size in $L$. We will merge all $N_\mu$ lists of size $\mu$ pairwise into $N_\mu/2$ lists of size $2\mu$. This increases the size of the smallest list in $L$ from $\mu$ to $2\mu$. We can repeat this process as long as the minimum list size $\mu$ is less than the maximum list size $N/\log^d N$. Eventually, all lists have size $N/\log^d N$. At that point, we will be able to apply the result from Part~1 of the proof, which deals precisely with the case that all lists have size $N/\log^d N$. This will show that $\E{F(L_{\rm final})}$ is negligible for the final list $L_{\rm final}$, obtained after all merging operations are done. In order to show that $\E{F(L)}$ is negligible for the list $L$ we start from, we will show that if $\E{F(L_{i+1})}$ is negligible for a list $L_{i+1}$ obtained \emph{after} a merging operation, then $\E{F(L_{i})}$ is also negligible for the list $L_i$ \emph{before} the merging operation. By induction, this will imply that since $\E{F(L_{\rm final})}$ is negigible, then $\E{F(L)}$ is also negligible for the original list $L$.

Thus, it suffices to show that if $\E{F(L)}$ is negligible after merging, then it was negligible before merging. This fact is the core of the proof, and involves several techniques. For now, we outline these techniques at a high level, and will provide more details when each technique is introduced.
Let $m_\mu = m/\mu$ be the number of superbuckets of size $\mu$. Let us regard lists of size $\mu$ as \emph{balls}, and superbuckets of size $\mu$ as \emph{bins}. Inserting the lists of size $\mu$ amounts to throwing $N_\mu$ balls into $m_\mu$ bins i.u.r. After merging, a list of size $2\mu$ is viewed as two connected balls. Each pair of connected balls is thrown i.u.r. into two adjacent bins (where the two adjacent bins correspond to one superbucket of size $2\mu$). When bins are inserted by pairs in that manner, one feature of the resulting distribution is that the bins with even indices (bins number 0, 2, 4, etc) must contain the same total number of balls as the bins with odd indices (bins number 1, 3, 5, etc).  When that property is satisfied, let us say that the bins are \emph{balanced}. To summarize what we have said so far: inserting merged lists will always produce balanced bins. On the other hand, if we insert lists before the merging step, there is no particular reason that the resulting bins should be balanced. The first main proof technique is to show the following: if we insert lists before the merging step, and condition the resulting distribution of bin occupancies on being \emph{balanced}, then the merging operation can only increase $\E{F(L)}$. This step relies on a convexity argument, and uses a special auxiliary operator $\diamond$. We leave a detailed discussion of those points for later, and continue to focus on the global outline of the proof.

Insofar as merging can only increase $\E{F(L)}$, we get what we want: if $\E{F(L)}$ is negligible after merging, then it was necessarily negligible before merging. However, to apply that argument, we need bins to be balanced. As mentioned earlier, there is no special reason that inserting $N_\mu$ balls into $m_\mu$ bins i.u.r. should result in balanced bins. This leads to the next proof technique, which is a stochastic dominance argument. Although the distribution obtained by throwing $N_\mu$ balls into $m_\mu$ bins i.u.r. is not balanced, we show that it is stochastically dominated by balanced distribution, namely the distribution obtained by throwing $N_\mu+\phi(\mu)$ balls into $m_\mu$ bins i.u.r. \emph{conditioned on being balanced}. Here, $\phi(\mu)$ is a carefully chosen small quantity. Intuitively, what happens is that although the original distribution may not be balanced, the difference $2\delta = |n_0 - n_1|$ between the number $n_0$ of balls in bins with even indices, and the number $n_1$ of balls in bins with odd indices, must be less than $\phi(\mu)$ (except with negligible probability). As a consequence, by adding less than $2\delta$ balls, we can ``correct'' the distribution into a balanced one, at the cost of slightly increasing the total number of balls. Since adding new balls can only increase the output of $\ov(\cdot)$, this new transformation has the desired property that if $\E{\max(\ov(\cdot)-R,0)}$ is negligible for the distribution at the output of the transformation, it was necessarily negligible before the transformation. On the other hand, because we add new balls, we need to be mindful that each merging increases the total number of balls in the system. However, we show that the total number of balls remains $O(N)$ throughout, which is enough for the proof to go through.

\subsubsection{Full Proof}

\begin{definition}
Let $a = (a_i)_{i \leq t}$ and $b = (b_i)_{i \leq t}$ be two vectors in $\N^t$. Then $a \diamond b$ denotes the following vector in $\N^{2t}$:
\[
a \diamond b = (a_1,b_1,a_2,b_2,\dots,a_t,b_t).
\]
\end{definition}

The notation $\diamond$ is extended in the usual way to combine two sets of vectors ($A \diamond B = \{a \diamond b : a \in A, b \in B\}$), and two distributions of vectors ($D_1 \diamond D_2 = a \diamond b$ where $a \gets D_1, b \gets D_2\}$).
The point of $\diamond$ is the next lemma, which is essentially a convexity argument.


\begin{lemma}
\label{lem:interleave}
Let $a = (a_i)_{i \leq t}$ and $b = (b_i)_{i \leq t}$ be two vectors in $\N^t$. We have:
\[
2F(a \diamond b) \leq F(a \diamond a) + F(b \diamond b).
\]
\end{lemma}

\begin{proof}
Let $a' \in \N^t$ be defined by $a'_i = \max(a_i - \tau,0)$, so that $a'_i$ is the number of overflowing elements in bucket $i$ for vector $a$. (Recall that $\tau$ is the threshold at which buckets are cut off.)
Define $b'$ in the same way.
Observe that $f:x \mapsto \max(x-R,0)$ is a convex function, which implies that for all $x, y$, $f(x/2+y/2) \leq (f(x) + f(y))/2$.
As a consequence:
\begin{align*}
2F(a \diamond b) &= 2\max\Big(\sum a'_i + \sum b'_i - R,0\Big)\\
&=2f\Big(\sum a'_i + \sum b'_i\Big)\\
&\leq f\Big(2\sum a'_i\Big) + f\Big(2\sum b'_i\Big)\\
&=F(a \diamond a) + F(b \diamond b).\qedhere
\end{align*}
\end{proof}


Given a load vector $b \in \N^{m}$, let $n_0(b) = \sum_{i:i \bmod 2 = 0} b_i$ (resp. $n_1(b) = \sum_{i:i \bmod 2 = 1} b_i$) be the total number of balls in bins with even (resp. odd) index.
Let $n(b) = n_0(b) + n_1(b)$ be the total number of balls.
Let $\delta(b) = \max(n_0(b), n_1(b)) - \lfloor n(b)/2\rfloor$.


Recall that $B_{n,m}[\delta = 0]$ denotes the distribution $B_{n,m}$ conditioned on the event $\delta = 0$, that is, the bins with even indices contain the same total number of balls as the bins with odd indices.
The proof of the following lemma is immediate.

\begin{lemma}
\label{lem:diamond}
For all even $n$, $m$:
\[
B_{n,m}[\delta = 0] = B_{n/2,m/2} \diamond B_{n/2,m/2}.
\]
\end{lemma}



Define:
\[
B'_{n,m,d} = B_{n,m}[\max(n_0,n_1) \leq n/2 + d].
\]

\begin{lemma}
\label{lem:statdist}
If $d = \Omega(\sqrt{n}\log \lambda)$, then
the statistical distance between $B_{n,m}$ and $B'_{n,m,d}$ is negligible.
\end{lemma}

\begin{proof}
Chernoff $\Rightarrow$ the probability that the condition that defines $B'_{n,m,d}$ is not satisfied in $B_{n,m}$ is negligible.
Further, if two distributions are identical conditioned on an event with negligible probability not happening, then their statistical distance is negligible (this is used implicitly all the time in game-hopping proofs).
\end{proof}


\begin{lemma}
\label{lem:stochdom}
For all $n$, $m$, $d$,
$B'_{n,m,d}$ is stochastically dominated by $B_{n+2d,m}[\delta = 0]$ (with respect to the product order on $\N^m$).
\end{lemma}

\begin{proof}
If we sample from $B'_{n,m,d}$, then add $n/2+2d - n_0$ (resp. $n/2+2d - n_1$) balls uniformly at random into buckets of even (resp. odd) indices, we obtain a sample from $B_{n+d,m}[\delta = 0]$.
Hence, there exists a suitable pairing of the two distributions.
\end{proof}


Let $\phi(\ell) = \sqrt{N/\ell}\log \lambda$. If $L$ is a multiset of list sizes, let $\mu = \min L$. Define $\iter(L)$ by removing all $N_\mu$ instances of $\mu$ from $L$, and adding instead
$N_\mu/2 + \phi(\mu)$ instances of size $2\mu$.

Let $N'_1 = N_1 + \sqrt{N}\log \lambda$. By induction, for $i$ in $\{1, \dots, \log N\}$ and $\ell = 2^i$, define:
\[
N'_{\ell} = N_\ell + \frac{N'_{\ell/2}}{2} + \sqrt{\frac{N}{\ell}}\log \lambda.
\]

\begin{lemma}
\label{lem:nprime}
For all $\ell \leq N/\log^2 \lambda$, $N'_\ell = O(N/\ell)$.
\end{lemma}

\begin{proof}
A straightforward induction gives:
\begin{flalign*}
&&N'_\ell &= N_\ell + \sum_{i = 0}^{\log \ell} \sqrt{\frac{N}{2^i}}\log \lambda&&\\
&&&= N_\ell + \sqrt{N}\log \lambda \sum_{i = 0}^{\log \ell} 2^{-i/2}&&\\
&&&= N_\ell + \sqrt{N}\log \lambda \cdot O\left(2^{-\frac{1}{2}\log \ell}\right)&&\\
&&&= N_\ell + O\left(\sqrt{\frac{N}{\ell}}\log \lambda\right)&&\\
&&&= N_\ell + O\left(\frac{N}{\ell}\right)&&\text{because $\log^2 \lambda \leq N/\ell$}\\
&&&= O\left(\frac{N}{\ell}\right).&&\hspace{2.7cm}\qedhere
\end{flalign*}
\end{proof}

\begin{lemma}
\label{lem:key}
Let $\mu = \min L$.
Assume $N_\mu$ is even.
If $\mu < N/\log^2 N$, then:
\[
\E{F(L)} \leq \E{F(\iter(L))} + \negl(\lambda).
\]
\end{lemma}

\begin{proof}
In the scope of this proof, $\mu$ is set to $\min L$.
Let $m_\mu = m/\mu$ be the number of superbuckets of size $\mu$.
Say that a superbucket is \emph{flat} iff all the buckets it contains have the same number of items.
Say that a vector of occupancies $b \in \N^m$ is \emph{$k$-flat} if all superbuckets of size $k$ are flat.

By construction of 1C, and that fact that $\mu = \min L$, after inserting lists in $L$, bucket occupancies are $\mu$-flat.
The load of a bucket is entirely determined by the number of items in the superbucket of size $\mu$ that contains it.
As a consequence, there is never a reason to consider superbuckets of size smaller than $\mu$.
For that reason, instead of working with $\N^m$, where each entry corresponds to the load of a bucket, we will work with $\N^{m_\mu}$, where each entry corresponds the load of a superbucket of size $\mu$, divided by $\mu$ (so that an entry is the load of one bucket within the superbucket). To avoid creating confusion about whether a ``bucket'' or ``items'' refers to the original occupancy vectors in $\N^m$, or the ones just introduced in $\N^{m_\mu}$, we reserve the term ``bucket'', ``superbucket'', and ``item'' to the former setting, so that the meaning of those terms in unchanged. When working in $N^{m_\mu}$, we use balls-and-bins terminology: $m_{\mu}$ is the number of \emph{bins}, and they are occupied by \emph{balls}. Thus, each \emph{bin} corresponds to a superbucket of size $\mu$, and each \emph{ball} corresponds to a list of $\mu$ items.

We now have all the tools to prove \Cref{lem:key}.
Let $L_\cap = L \cap \iter(L)$ be the lists common to $L$ and $\iter(L)$.
Observe that the order lists are inserted by 1C does not matter, hence we are free to assume lists in $L_\cap$ are inserted first.

Let $a \in \N^{m_\mu}$ denote an arbitrary load vector obtained after inserting the lists in $L_\cap$.
Recall that lists in $L_\cap$ are multiples of $2\mu$, so the load vector after inserting the lists is $2\mu$-flat.
If follows that $a$ may be written in the form $a = a' \diamond a'$ for some $a' \in \N^{m_\mu/2}$.
Let us denote by $E_a$ the event that the outcome of inserting $L_\cap$ is equal to $a$.

We want to prove $\E{F(L)} \leq \E{F(\iter(L))} + \negl(\lambda)$.
By \Cref{lem:condexp}, it suffices to prove the inequality when conditioning on $E_a$, for every possible $a$.

Given $E_a$, all that remains to do to compute $D(L)$ is to insert $N_\mu$ lists of length $\mu$.
In consequence, we have that $D(L)$ conditioned on $E_a$ is equal to $a + B_{N_\mu,m_\mu}$.
The lemma can then be established as follows.

In the computation, we multiply the output of $\ov$ by $\mu$, to reflect the fact that each ball in $N^{m_\mu}$ represents a list of $\mu$ items.
\begingroup
\allowdisplaybreaks
\begin{flalign*}
&&\E{F(L) : E_a} &= \E{\max(\mu\cdot\ov(D(L)-R,0) : E_a}&&\\
&&&= \E{\max(\mu\cdot\ov(a + X-R,0)}&&\\
&&&\quad\text{where $X \hookleftarrow B_{N_\mu,m_\mu}$}&&\\
&&&\leq \E{\max(\mu\cdot\ov(a + X'-R,0)} + N \negl(\lambda)&&\\
&&&\quad\text{where $X' \hookleftarrow B'_{N_\mu,m_\mu,\phi(\mu)}$}&&\text{by \Cref{lem:statdist}}\footnotemark.\\
&&&\leq \E{\max(\mu\cdot\ov(a + Y-R,0)} + \negl(\lambda)&&\\
&&&\quad\text{where $Y \hookleftarrow B_{N_\mu + 2\phi(\mu),m_\mu}[\delta=0]$}&&\text{by \Cref{lem:stochdom}}\\
&&&= \E{\max(\mu\cdot\ov(a + Y^1 \diamond Y^2) - R,0)} + \negl(\lambda)&&\\
&&&\quad\text{where $Y^1, Y^2 \hookleftarrow B_{(N_\mu + 2\phi(\mu))/2,m_\mu/2}$}&&\text{by \Cref{lem:diamond}}\\
&&&\leq \E{\max(\mu\cdot\ov(a + Y^1 \diamond Y^1) - R,0)}/2&&\\
&&&\quad + \E{\max(\mu\cdot\ov(a + Y^2 \diamond Y^2) - R,0)}/2 + \negl(\lambda)&&\text{by \Cref{lem:interleave}}\\
&&&=\E{\max(\mu\cdot\ov(a + Y^1 \diamond Y^1) - R,0)} + \negl(\lambda)&&\\
&&&=\E{\max(\mu\cdot\ov((a' + Y^1) \diamond (a' + Y^1)) - R,0)} + \negl(\lambda)&&\\
&&&=\E{\max(2\mu\cdot\ov(a' + Y^1) - R,0)} + \negl(\lambda)&&\\
&&&=\E{F(\iter(L)) : E_a} + \negl(\lambda).&&\hspace{1.4cm}\qedhere
\end{flalign*}
\endgroup
\end{proof}

\footnotetext{The fact that the condition $d = \Omega(\sqrt{n}\log \lambda)$ from \Cref{lem:statdist} is satisfied follows from \Cref{lem:nprime}.}

If we start from an arbitrary $L$, by applying \Cref{lem:key} and computing $\iter(L)$ as in the statement of the lemma, we strictly increase the minimum size of the list.
Eventually, all lists have size $N/\log^2 N$, while the total number of items remains $O(N)$ (\Cref{lem:nprime}). Hence, the analysis from Part~1 applies, and we are done.


\end{document}